\documentclass[a4paper,11pt]{article}
\usepackage{jheppub}
\makeatletter
\gdef\@fpheader{}
\makeatother
\usepackage{lineno}
\usepackage{xcolor}
\usepackage{slashed} 
\usepackage{amsmath}
\DeclareUnicodeCharacter{2113}{\ensuremath{\ell}}
\allowdisplaybreaks

\title{\boldmath Hadronic photon correction to  $\gamma^{\ast} \gamma \to f_{2}(1270)$ at next-to-leading order}

\author{Bing-Xin Liu and Shen-Qiang Ren}
\affiliation{School of Physics, Nankai University, 300071 Tianjin, China}

\emailAdd{lbx@mail.nankai.edu.cn, aersq@mail.nankai.edu.cn}

\abstract{Within the framework of light-cone sum rules, we calculate the hadronic photon corrections to the $\gamma^*\gamma \to f_2(1270)$ transition form factors induced by the leading-twist photon distribution amplitude of the real photon. We establish the factorization formula for the vacuum-to-photon correlation function at next-to-leading order in $\alpha_s$, and extract the perturbative hard matching coefficients by applying the method of regions. The parametrically large logarithms appearing in the hard functions are resummed to next-to-leading logarithmic accuracy by solving the two-loop evolution equation for the corresponding light-ray tensor operator. Combining the resulting light-cone sum rules with the known leading-power contributions from QCD collinear factorization, we provide updated theoretical predictions for the three helicity form factors $T_0(Q^2)$, $T_1(Q^2)$ and $T_2(Q^2)$, including an estimate of the theoretical uncertainties.}

\begin{document}
\maketitle
\flushbottom

\section{Introduction}
\label{sec:intro}
Among the various hadronic states, tensor mesons with $J^P=2^+$ have attracted significant attention due to their distinctive spin-parity quantum numbers. In contrast to pseudoscalar and vector mesons, tensor mesons can not be produced by local vector or axial-vector currents. Moreover, the different polarizations of the tensor mesons in $B$-meson decays can shed light on the helicity structure of the electroweak interactions~\cite{BaBar:2008ozy, BaBar:2008lan, Cheng:2010yd,Kim:2013cpa}. The two-quark light-cone distribution amplitudes of tensor mesons, including twist-2 and twist-3 components, have been systematically studied, and the relevant decay constants have been estimated using QCD sum rule techniques~\cite{Cheng:2010hn}.

As the lightest tensor meson with quantum numbers $J^{PC}=2^{++}$, the $f_2(1270)$ serves as a pivotal object for exploring hadronic structure, tensor glueball, quark–glueball mixing, and nonperturbative QCD effects~\cite{Cheng:2010hn,Braun:2000cs,Braun:2016tsk,Giacosa:2005bw,Lebiedowicz:2020qnz}. The $f_2(1270)$ meson was first observed in pion-proton scattering experiments, and has since been measured in $e^+e^-$ collisions, radiative $J/\psi$ decays, and $B$-meson decays, e.g., $J/\psi \to \omega f_2(1270)$, $B^+ \to K^+ f_2(1270)$~\cite{Belle:2005rpz,BaBar:2008lpx,BESIII:2015rug}. The Belle collaboration measured the $\gamma^* \gamma \to f_2(1270)$ transition form factors via single-tag two-photon process for $Q^2$ from a few to 30 $\mathrm{GeV}^2$~\cite{Belle:2015oin}. This electromagnetically induced hard exclusive process provides an important platform for probing the quark and gluon components of tensor mesons, investigating their internal structure, and testing the QCD scaling behavior. The $f_2$ meson transition form factors were discussed in~\cite{Schuler:1997yw,Hoferichter:2020lap}. While the leading-power contribution to the $\gamma^*\gamma \to f_2(1270)$ transition form factors has been systematically investigated within the QCD factorization framework~\cite{Braun:2000cs,Braun:2016tsk}, the relevant subleading-power effects remain to be further explored. To complement these theoretical predictions and facilitate a more comprehensive comparison with experimental measurements, our work incorporates the hadronic photon corrections to $\gamma^* \gamma \to f_2(1270)$ into the existing theoretical framework. Similar corrections have been shown to have a significant impact in the two-photon process $\gamma^* \gamma \to \pi^0$~\cite{Wang:2017ijn}.

Motivated by these earlier studies, the primary objective of this work is to systematically evaluate the $\gamma^*\gamma \to f_2(1270)$ form factors by calculating the subleading power corrections induced by the hadronic component of the energetic photon at next-to-leading-order (NLO) accuracy in the strong coupling constant. To separate the hard scattering from the nonperturbative dynamics and systematically extract these power-suppressed contributions, our theoretical calculations are carried out employing the technique of light-cone sum rules (LCSRs)~\cite{Braun:1997kw,Colangelo:2000dp} with the two-particle photon distribution amplitudes (DAs). Based on QCD factorization and dispersion relations, this method employs the Borel transformation to suppress the contributions from excited states and the continuum, thereby naturally avoiding the endpoint divergences. In recent years, this method has been successfully applied to evaluate the next-to-leading-power (NLP) and higher-order power corrections in various related processes, including those associated with hadronic photon effects, such as the $\gamma^* \gamma \to \pi^0$ form factor~\cite{Agaev:2010aq,Mikhailov:2016klg,Wang:2017ijn,Shen:2019zvh,Stefanis:2020rnd}, $\gamma^* \gamma \to \eta^{\prime}$~\cite{Agaev:2014wna,Braun:2025xpc}, and the $B$-meson decays (see for instance~\cite{Sun:2010nv,Zhong:2011jf,Wang:2017jow,Wang:2018wfj,Emmerich:2018rug,Gao:2019lta,Beneke:2020fot,Li:2020rcg,Gao:2021sav,Cui:2022zwm,Huang:2025jsa,Di:2025hdu,Gao:2024vql}). A recent comprehensive account of the LCSR framework for b-quark decays can be found in~\cite{Khodjamirian:2023wol}. In this work, we will establish an NLO factorization formula for the NLP contributions by calculating the one-loop QCD corrections to the vacuum-to-photon correlation function and  applying the method of regions~\cite{Beneke:1997zp} to extract the hard matching coefficients. We will also achieve a next-to-leading logarithmic (NLL) resummation of the parametrically large logarithms $\ln(\mu^2/Q^2)$ by solving the two-loop renormalization-group evolution equation for the light-ray tensor operator. Combining our newly obtained hadronic photon effects with the known leading-power (LP) contributions from collinear factorization, we will deliver up-to-date theoretical predictions for the form factors $T_0{(Q^2)}$, $T_1{(Q^2)}$ and $T_2{(Q^2)}$.

The outline of this paper is as follows: in Section~\ref{sect:Production f_2} we present the hadronic real photon-to-vacuum correlation function with an electromagnetic current $j_{\mu}^{\mathrm{em}}$ and a tensor meson interpolating current $j_{\rho\sigma\alpha}$, and the definitions of the form factors for $\gamma^*\gamma \to f_2(1270)$. In Section~\ref{subleading-power}, we establish the tree-level and one-loop factorization formulas for the vacuum-to-photon correlation function, extract the perturbative hard matching coefficients, and derive the corresponding NLL resummation improved LCSRs for the hadronic photon contributions to the $\gamma^*\gamma\to f_2(1270)$ transition form factors. Combining the resulting hadronic photon contributions with the known LP contributions, we present a phenomenological analysis of the three form factors in Section~\ref{Numerical analysis}. A summary of our main observations and concluding remarks is presented in Section~\ref{Conclusion}. We present the spectral representations used to analytically continue the QCD correlation functions in the variable $p^2$ to the physical region in Appendix~\ref{app:A}, and summarize the existing LP results for the transition form factors in Appendix~\ref{app:B}.

\section{Production of $f_2(1270)$ in two-photon reactions}
\label{sect:Production f_2}
For the reaction $\gamma(p^{\prime}) \, \gamma^*(q) \to f_{2}(1270)$ with one real photon ($p^{\prime \, 2}=0$) and one virtual photon ($q^{2}=-Q^{2}$),  the produced tensor meson  $f_2(1270)$ has quantum numbers  $J^{PC}=2^{++}$ and $I^G=0^{+}$. In the SU(2)-flavor limit, the dominant nonstrange component of $f_2(1270)$ is taken to be $1/\sqrt{2}\left [\bar{u}u+\bar{d}d \right ]$. To evaluate the power-suppressed contribution of the hadronic photon effect to the tensor meson form factors at NLO with the LCSR approach, we consider the following vacuum-to-photon correlation function involving an electromagnetic current carrying a four-momentum $q_{\mu}$ and an interpolating current $j_{\rho\sigma\alpha}$ \cite{Wang_2011, Cheng:2010hn} for the tensor meson
\begin{align}
\label{eq:2.1}
G_{\mu\rho\sigma\alpha}(p^{\prime}, q) = \int d^4 x \, e^{-i \, q \cdot x}  \,
\langle 0 | {\rm T} \left \{ j_{\mu}^{\text{em}}(x), j_{\rho\sigma\alpha}(0) \right \} | \gamma(p^{\prime},\epsilon) \rangle  \,,
\end{align}
where $p=p'+q$ denotes the momentum flowing through the tensor-meson interpolating current, $p'$ is the four-momentum of the on-shell photon, and $q$ is the transfer momentum. The relevant currents are defined as
\begin{align}
j_{\mu}^{\text{em}}=\sum_q \, g_\mathrm{em} \, Q_q \, \bar{q} \,  \gamma_\mu \, q \,, \qquad
j_{\rho\sigma\alpha}=\frac{1}{\sqrt{2}} \left( \bar{u} \, \sigma_{\rho\sigma} \, i \, \overset{\leftrightarrow}{\operatorname*{\mathit{D}}}_{\alpha} \, u + \bar{d} \, \sigma_{\rho\sigma} \, i \, \overset{\leftrightarrow}{\operatorname*{\mathit{D}}}_{\alpha} \, d \right)\,.
\end{align}
The covariant
derivative is defined as $\overset{\leftrightarrow}{D}_\alpha =\overset{\rightarrow}{D}_\alpha - \overset{\leftarrow}{D}_\alpha$, with $\overset{\rightarrow}{D}_\alpha = \overset{\rightarrow}{\partial}_\alpha + i g_s A^a_\alpha \lambda^a/2$ and $\overset{\leftarrow}{D}_\alpha = \overset{\leftarrow}{\partial}_\alpha - i g_s A^a_\alpha \lambda^a/2$. The decay constant of $f_2$ is defined by the matrix element of the following operator
\begin{align}
\label{eq:2.3}
\langle f_2(p,\lambda)|j_{\rho\sigma\alpha}|0\rangle=-if_{f_2}^{\perp}m_{f_2}(\epsilon_{\rho\alpha}^{(\lambda)*}p_{\sigma}-\epsilon_{\sigma\alpha}^{(\lambda)*}p_{\rho}),
\end{align}
where $m_{f_2} = 1270~\mathrm{MeV}$ is the mass of $f_2$, and the symmetric and traceless polarization tensor satisfies ${\epsilon}_{\mu\nu}^{(\lambda)}p^{\nu}=0$. The polarization sum formula is
\begin{align}
\sum_{\lambda} \epsilon_{\mu\nu}^{(\lambda)} \epsilon_{\rho\sigma}^{(\lambda)*}  = \frac{1}{2} M_{\mu\rho} M_{\nu\sigma} + \frac{1}{2} M_{\mu\sigma} M_{\nu\rho} - \frac{1}{3} M_{\mu\nu} M_{\rho\sigma} \, ,
\end{align}
where $M_{\mu\nu}=g_{\mu\nu}-p_{\mu}p_{\nu}/m_{f_2}^{2}$.

Following Refs.~\cite{Braun:2000cs,Braun:2016tsk}, the amplitude for the process $\gamma^*\gamma\to f_2(1270)$ can be related to the matrix element
\begin{align}
T_{\mu\nu}=i \, \int d^4 y \, e^{-i \, q \cdot y}  \,
\langle f_2(p,\lambda) | {\rm T} \left \{ j_{\mu}^{\text{em}}(y), j_{\nu}^{\text{em}}(0) \right \} | 0 \rangle  \, .
\end{align}
The hadronic matrix element $T^{\mu\nu}$ can be decomposed in terms of three Lorentz structures
\begin{align}
\label{eq:2.6}
T^{\mu\nu}=T_0^{\mu\nu}+T_1^{\mu\nu}+T_2^{\mu\nu},
\end{align}
with
\begin{align}
T_{0}^{\mu \nu}&={\epsilon}_{\beta \gamma}^{(\lambda) *}\left(-g_{\perp}^{\mu \nu}\right)\left(q-p'\right)^{\beta}\left(q-p'\right)^{\gamma} \frac{m^{2}_{f_2}}{\left(2\,  q \, p'\right)^{2}} \, T_{0}\left(Q^{2}\right),  \nonumber\\[0.5em]
T_{1}^{\mu \nu}&={\epsilon}_{\beta \gamma}^{(\lambda) *}\left(-g_{\perp}^{\beta \nu}\right)\left(q-p'\right)^{\gamma}\left[q^{\mu}-p'^{\mu} \frac{q^{2}}{\left(q \, p'\right)}\right] \frac{m^{2}_{f_2}}{\left(2 \, q \,  p'\right)^{2}} \, T_{1}\left(Q^{2}\right), \nonumber \\[0.5em]
T_{2}^{\mu \nu}&= {\epsilon}_{\beta \gamma}^{(\lambda) *}\left[g_{\perp}^{\beta \mu} g_{\perp}^{\gamma \nu}-\frac{1}{2} g_{\perp}^{\mu \nu} \frac{m^{2}_{f_2}}{\left(2 \, q \, p'\right)^{2}}\left(q -p'\right)^{\beta}\left(q -p'\right)^{\gamma}\right] \, T_{2}\left(Q^{2}\right),
\end{align}
where the transverse metric tensor $g_{\perp}^{\mu\nu}$ is defined as
\begin{align}
g_{\perp}^{\mu\nu} = g^{\mu\nu} - \frac{1}{(q \, p')} (q^\mu p'^\nu + q^\nu p'^\mu) + \frac{q^2}{(q \, p')^2} \,  p'^\mu p'^\nu \,.
\end{align}

For the evaluation of the hadronic photon contribution, we introduce the following light-cone projected components of the tensor-meson interpolating current $j_{\rho\sigma\alpha}$ to isolate the contributions associated with the helicity form factors $T_0$, $T_1$, and $T_2$:
\begin{align}
j_{0,\rho\sigma\alpha}&= \bar{q} \,\frac{i}{2} \,  \frac{\bar{n}_{\rho}}{2} \,\frac{n_{\sigma}}{2} \, \frac{n_{\alpha}}{2} \, ( \slashed{n} \slashed{\bar{n}}- \slashed{\bar{n}}  \slashed{n}) \, \bar{n} \cdot \overset{\leftrightarrow}{\operatorname*{\mathit{D}}} \, q \, , \nonumber \\[0.5em]
j_{1,\rho\sigma\alpha}&=\bar{q} \, \frac{i}{2} \, \frac{\bar{n}_{\sigma}}{2} \, \frac{n_{\alpha}}{2} \, (\gamma^{\perp}_{\rho} \slashed{n}-\slashed{n}\gamma^{\perp}_{\rho}) \, \bar{n} \cdot \overset{\leftrightarrow}{\operatorname*{\mathit{D}}} \, q \, , \nonumber \\[0.5em]
j_{2,\rho\sigma\alpha}&=\bar{q} \, \frac{i}{2} \, \frac{\bar{n}_{\sigma}}{2} \,(\gamma^{\perp}_{\rho} \slashed{n}-\slashed{n}\gamma^{\perp}_{\rho}) \, \overset{\leftrightarrow}{\operatorname*{\mathit{D}}}^{\perp}_{\alpha} \, q \, .
\end{align}
The two light-like vectors $n$ and $\bar{n}$ satisfy the conditions $n^2=\bar{n}^2=0$ and $n \cdot \bar{n}=2$. We choose $\bar{n}$ along the real-photon momentum $p^{\prime}$, while $n$ denotes the conjugate light-cone direction. The transverse Dirac matrices are defined accordingly as $\gamma_{\mu}^{\perp} = \gamma_{\mu} -  \slashed{n} \bar{n}_{\mu} /2-  \slashed{\bar{n}} n_{\mu}/2$. we further adopt the following power counting scheme in the regime of large momentum transfer where $Q^2\gg\Lambda_{\text{QCD}}^2 $
\begin{align}
\label{eq:2.8}
| n \cdot p | \sim \bar n \cdot p \sim n \cdot p^{\prime} \sim {\cal O}(\sqrt{Q^2}) \,.
\end{align}
This kinematic setup provides a systematic organization of the correlation function in the large-$Q^2$ expansion and facilitates the factorization analysis of the hadronic photon contribution and the subsequent construction of the corresponding LCSRs for the transition form factors.
\begin{figure}[htbp]
\centering
\includegraphics[width=0.8\textwidth]{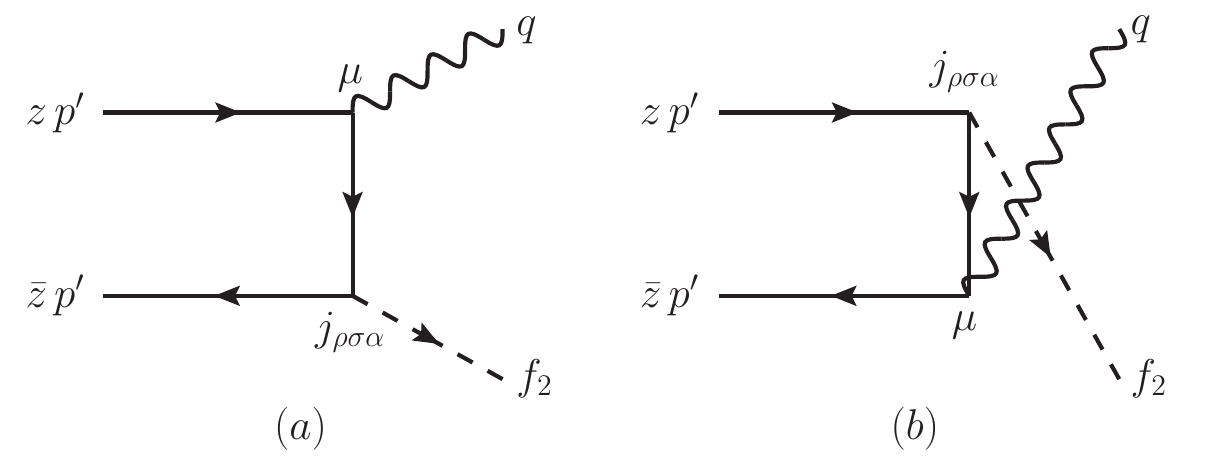}
\caption{Diagrammatical representation of the correlation function  $\Pi_{\mu\rho\sigma\alpha }$ at tree level. \label{fig:1}} 
\end{figure}

\section{The subleading-power corrections from photon DAs}
\label{subleading-power}
In this section, we investigate the leading-twist hadronic photon contributions to the $\gamma^*\gamma\to f_2(1270)$ transition form factors within the LCSR framework. The factorization formulas for the vacuum-to-photon correlation function are established at tree level and extended to one loop by calculating the QCD radiative corrections and extracting the perturbative hard matching coefficients. The large logarithms are resummed to NLL accuracy using the two-loop evolution equation of the leading-twist photon DA. By matching the QCD correlation functions onto their hadronic dispersion representations and performing the Borel transformation and continuum subtraction, we obtain the NLL LCSRs for the three form factors $T_i(Q^2)$, with $(i=0,1,2)$.

\subsection{The hadronic photon effect at tree level}
\label{sect:tree level}
To construct the  QCD factorization formulas for the correlation function \eqref{eq:2.1} at tree level, we consider the following four-point QCD amplitude
\begin{align}
\Pi _{i,\mu\rho\sigma\alpha} = \int d^4 x \, e^{-i \, q \cdot x}  \,
\langle 0 | {\rm T} \left \{ j_{\mu}^{\text{em}}(x), j_{i,\rho\sigma\alpha}(0) \right \} | q(k_1) \,\bar{q}(k_2) \rangle \,,  
\end{align}
 where $k_1=z \,p'$, $k_2=\bar{z} \,p'$, $z$ indicates the momentum fraction carried by the collinear quark of the real photon and $\bar z = 1-z $. The partonic amplitude for diagram (a) in figure~\ref{fig:1} reads
 \begin{align}
\Pi _{\mu\rho\sigma\alpha}^{(0,a)}=\frac{i \, g_{\text{em}} \, Q_q \, (p-2k_2)_\alpha}{(p-k_2)^2 + i0}  \, \bar{v}(k_2) \, \sigma_{\rho\sigma} \, (\slashed{p} - \slashed{k}_2) \, \gamma_\mu \, u(k_1) \,.
\end{align}
Evaluating diagram (a) in figure~\ref{fig:1} and the corresponding diagram with the electromagnetic current attached to the antiquark line, which is obtained by interchanging $z \leftrightarrow \bar{z}$, leads to 
\begin{align}
\Pi _{0,\mu\rho\sigma\alpha}^{\,(0)} &=-\frac{  g_{\text{em}} \, \bar{n} \cdot p }{\sqrt{2} \, Q^{2}}\, (p-2k_{2})_{\alpha} \, \frac{\bar{n}_{\rho}}{2} \, \frac{n_{\sigma}}{2} \, \left [ \frac{1}{z \, r+\bar{z}}+ \frac{1}{\bar{z} \, r+z} \right ] \sum_{q=u,d}   Q_{q} \,\bar{v}(k_2) \, \slashed{n} \, {\gamma}_{\mu}^{\perp} \, u(k_1) \nonumber \\
 &= -\frac{ g_{\text{em}} \, \bar{n} \cdot p }{\sqrt{2} \, Q^{2}}\, (p-2k_{2})_{\alpha} \, \frac{\bar{n}_{\rho}}{2} \, \frac{n_{\sigma}}{2} \, \sum_{q=u,d}  Q_{q} \,\left [ \frac{1}{z^{\prime} \, r+\bar{z}^{\prime}}+ \frac{1}{\bar{z}^{\prime} \, r+z^{\prime}} \right ]* \langle O_{\mu}(z, z^{\prime}) \rangle ^{(0)} \, , \nonumber \\[0.5em]
 \Pi _{1,\mu\rho\sigma\alpha}^{\,(0)} &=
\frac{2 \,  g_{\text{em}} \,  n\cdot (p-k_{2})}{\sqrt{2} \, Q^{2}}(p-2k_{2})_{\alpha} \, \frac{\bar{n}_{\mu}}{2} \, \frac{\bar{n}_{\sigma}}{2}\left [ \frac{1}{z \, r+\bar{z}}+ \frac{1}{\bar{z} \, r+z} \right ] \, \sum_{q=u,d} Q_{q} \,  \bar{v}(k_2) \, \slashed{n} \,  {\gamma}_{\rho}^{\perp} \, u(k_1) \nonumber \\
&=  \frac{2 \,  g_{\text{em}} \,  n\cdot (p-k_{2})}{\sqrt{2} \, Q^{2}}(p-2k_{2})_{\alpha} \, \frac{\bar{n}_{\mu}}{2} \, \frac{\bar{n}_{\sigma}}{2} \sum_{q=u,d}Q_{q} \, \left [\frac{1}{z^{\prime} \, r+\bar{z}^{\prime}}+ \frac{1}{\bar{z}^{\prime} \, r+z^{\prime}}\right ]* \langle O_{\rho}(z, z^{\prime}) \rangle ^{(0)} \, , \nonumber \\[0.5em]
\Pi _{2,\mu\rho\sigma\alpha}^{\,(0)} &=0 \, ,  
\end{align}
where $r=-p^{2}/Q^{2}$, $\bar{r}=1-r$ and  the convolution integral of $z^{\prime}$ is represented by an asterisk. The partonic matrix element of the soft-collinear effective theory (SCET) operator $\langle O_{\mu}(z, z^{\prime}) \rangle ^{(0)}$ at tree level is given by
\begin{align}
\langle O_{\mu}(z, z^{\prime})\rangle  = \langle 0 | O_{\mu}(z^{\prime}) | q(k_1) \, \bar{q}(k_2) \rangle = \bar{\chi}(k_2) \, \slashed{n} \, {\gamma}_{\mu}^{\perp} \, \chi(k_1) \, \delta(z - z^{\prime}) + \mathcal{O}(\alpha_s) \, ,
\end{align}
the momentum-space definition of the SCET operator $ O_{\mu}(z^{\prime})$ reads
\begin{align}
 O_{\mu}(z^{\prime})=\frac{n \cdot p^{\prime}}{2 \, \pi}\int d \tau \, e^{i \, z^{\prime} \, \tau \, n \cdot p^{\prime} } \, \bar{\chi}(0) \, W_{\bar{c}}(0,\tau n)\, \slashed{n} \, {\gamma}_{\mu}^{\perp}  \, \chi(\tau n)\,,
\end{align}
where the Wilson line is
\begin{align}
W_{\bar{c}}(0,\tau n) = \mathrm{P}\left\{\mathrm{Exp}\left[-i\,g_s\int_0^\tau d\lambda\, n\cdot A_{\bar{c}}(\lambda\, n)\right]\right\}\, .
\end{align}
Using the leading-twist photon DA defined in~\cite{Ball_2003}
\begin{align}
&\langle0|\,\bar{q}(0)\,\slashed{n} \, \gamma^{\perp}_{\mu}\,[0,x]\,q(x)\,|\gamma(p^{\prime},\epsilon)\rangle \nonumber \\
&=- g_{\text{em}}Q_{q}  \, \chi(\mu) \, \langle\bar{q}q\rangle(\mu) \, n \cdot p^{\prime} \, \epsilon^{\perp}_{\mu}\int_{0}^{1}dz \, e^{-i z \, p^{\prime} \cdot x}\phi_{\gamma}(z,\mu) \, ,
\end{align}
we can write down the tree-level factorization formulas for the correlation function $G_{\mu\rho\sigma\alpha}(p^{\prime},q)$ as
\begin{align}
G_{0,\mu\rho\sigma\alpha}^{(0)}(p^{2},Q^2)=& \sqrt{2}\, g_{\text{em}}^2 \, \sum_{q=u,d} Q^2_{q} \, \frac{n_{\alpha}}{2} \, \frac{\bar{n}_{\rho}}{2} \, \frac{n_{\sigma}}{2} \, \chi(\mu) \, \langle\bar{q}q\rangle(\mu) \, \epsilon_{\mu}^{\perp}(p^{\prime}) \nonumber \\
&\times \int_0^1 dz \, \frac{(\bar{n} \cdot p)^2 \, n \cdot p^{\prime}}{z \, p^2-\bar{z} \, Q^2 +i0} \, \phi_{\gamma}(z,\mu) + \mathcal{O}(\alpha_s) \, , \nonumber \\[0.5em]
G_{1,\mu\rho\sigma\alpha}^{(0)}(p^{2},Q^2)=& -2\sqrt{2}\, g_{\text{em}}^2 \, \sum_{q=u,d} Q^2_{q} \, \frac{n_{\alpha}}{2} \, \frac{\bar{n}_{\sigma}}{2} \, \frac{\bar{n}_{\mu}}{2} \, \chi(\mu) \, \langle\bar{q}q\rangle(\mu) \, \epsilon_{\rho}^{\perp}(p^{\prime}) \nonumber \\
&\times \int_0^1 dz \, n \cdot p^{\prime} \, \phi_{\gamma}(z,\mu) + \mathcal{O}(\alpha_s) \, , \nonumber \\[0.5em]
G_{2,\mu\rho\sigma\alpha}^{(0)}(p^{2},Q^2)= & \,0 \,.
\end{align}

Using the definition of the $f_2$ meson decay constant in \eqref{eq:2.3} and the parametrization of the $\gamma^* \gamma \to f_2(1270)$ 
transition form factors $T_i(Q^2)$ in \eqref{eq:2.6}, we obtain the hadronic dispersion relation for $G_{\mu\rho\sigma\alpha}(p^{\prime},q)$ as 
\begin{align}
\label{eq:3.8}
G_{\mu\rho\sigma\alpha}(p^{2},Q^2)= &\frac{f_{f_2}^{\perp} \, m_{f_2}}{m_{f_2}^2-p^2-i0} \, \epsilon^{\nu} \, (p^{\prime}) \, \sum_\lambda (\epsilon_{\rho\alpha}^{(\lambda)} \, p_{\sigma}-\epsilon_{\sigma\alpha}^{(\lambda)} \, p_{\rho}) \, T_{\mu\nu}^{(\lambda)} + \int_{s_0}^{\infty}ds \, \frac{\rho^{h}_{\mu\rho\sigma\alpha}(s,Q^2)}{s-p^2-i0}\,,
\end{align}
where $\rho^{h}_{\mu\rho\sigma\alpha}(s,Q^2)$ represents the hadronic spectral density for excited and continuum states with $f_2(1270)$  quantum numbers. By performing the Borel transformation to suppress the contributions from excited and continuum states, and employing quark-hadron duality to effectively model the continuum contribution, we obtain the tree-level LCSR expressions for the hadronic photon contribution to the $\gamma^*\gamma\to f_2(1270)$ transition form factors
\begin{align}
T_0^{(0)}(Q^2) &= -\frac{3 }{2\sqrt{2}} \frac{Q^2 \, g_{\text{em}}^2 \, } {f_{f_2}^{\perp} \, m_{f_2} }\sum_{q=u,d}Q_q^2 \, \chi(\mu) \langle \bar{q} q \rangle (\mu) \int_{z_0}^{1} \frac{dz}{z^2} \, \exp\left[ -\frac{\bar{z} \, Q^2-z \, m_{f_2}^2}{z \, M^2} \right]  \phi_{\gamma}(z, \mu)+ \mathcal{O}(\alpha_s) \, , \nonumber \\[0.5em]
T_1^{(0)}(Q^2) &= 0, \qquad T_2^{(0)}(Q^2) = 0,
\end{align}
where $z_0= Q^2/(s_0+Q^2)$, $M^2$ and $s_0$ denote the Borel parameter and the effective threshold, respectively. For $T_1^{(0)}$, the tree-level partonic correlation function is nonvanishing. However, this contribution does not generate a pole term in the $f_2$ channel. For the interpolating current $j_{1,\rho\sigma\alpha}$, the collinear equations of motion imply that only the $n$-projected component of the internal propagator numerator $(\slashed p-\slashed k_2)$ contributes. This numerator factor cancels the pole-producing factor in the propagator denominator. Therefore the nonzero partonic contribution to $G_{1,\mu\rho\sigma\alpha}^{(0)}$ contains no $f_2$-channel pole and does not contribute to the LCSR for $T_1^{(0)}$. The vanishing of $T_2^{(0)}$ at tree level can be understood from the fact that the transverse derivative in $j_{2,\rho\sigma\alpha}$ projects out $(p-2k_2)_\alpha^\perp$, which vanishes because the external momenta contain no transverse components.

Employing the power-counting scheme $s_0 \sim M^2 \sim \mathcal{O}(\Lambda^2)$ and $\bar{z}_0 \sim \mathcal{O}(\Lambda^2 / Q^2)$ for the sum rule parameters, and utilizing the leading-power form factor results presented in \cite{Braun:2016tsk} (the explicit expressions for $T_i^{\text{LP}}$ are summarized in Appendix~\ref{app:B}), we can obtain the scaling behavior of the hadronic photon effect at large $Q^2$
\begin{align}
\label{eq:3.10}
\frac{T_{0}^{\text{NLP}}(Q^2)}{T_{0}^{\text{LP}}(Q^2)} \sim \mathcal{O}\left(\frac{\Lambda^2}{Q^2}\right).
\end{align}
This ratio shows that in the asymptotic hard-scattering regime $Q^2 \gg \ \Lambda^2$, the hadronic photon contribution to $T_0$ is power suppressed relative to the LP contribution, scaling as $ \Lambda^2/Q^2$ in the large-$Q^2$ limit.
\begin{figure}[htbp]
\centering
\includegraphics[width=1\textwidth]{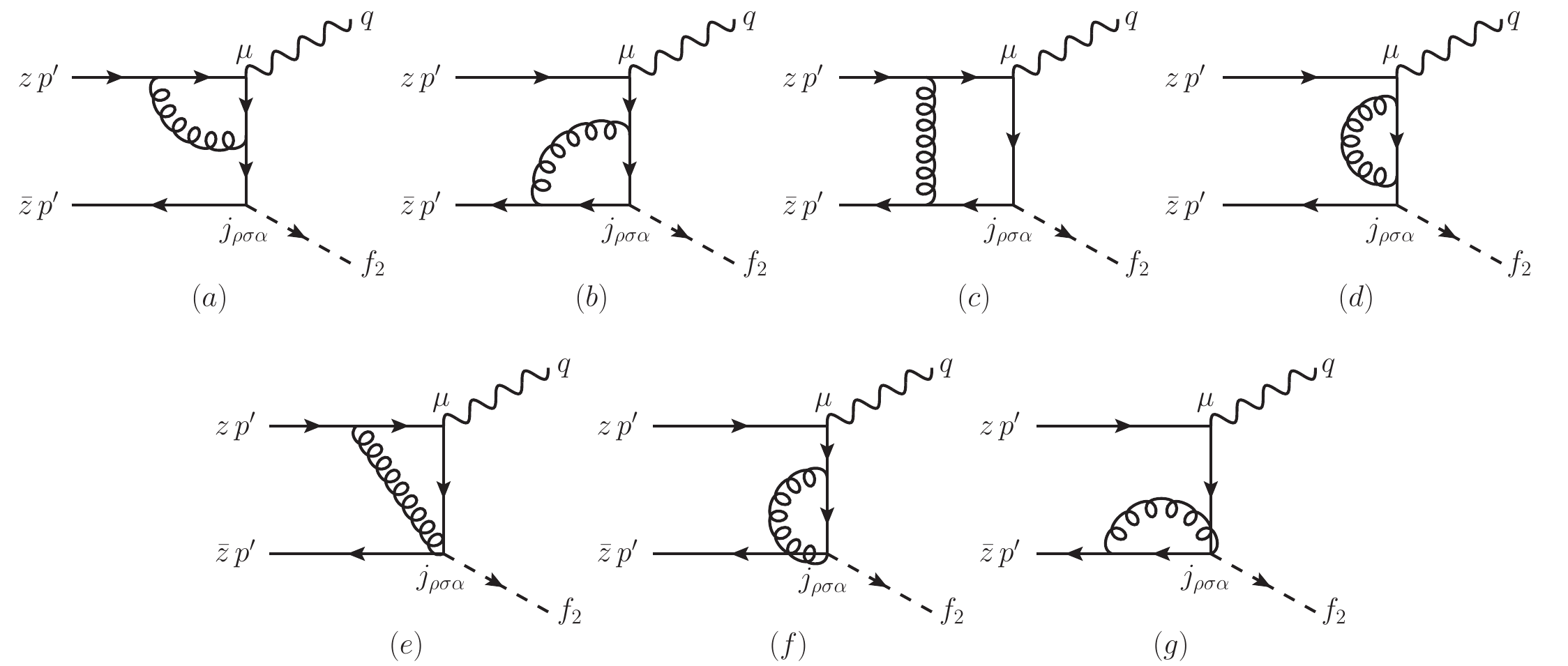}
\caption{Diagrammatical representation of the correlation function  $\Pi_{\mu\rho\sigma\alpha }$ at one loop. The corresponding symmetric diagrams obtained by exchanging the electromagnetic current and the tensor meson interpolating current are not presented here.\label{fig:2}}
\end{figure}

\subsection{The hadronic photon effect at one loop}
\label{one loop}
In this section, we calculate the one-loop radiative corrections to the hadronic photon contributions to the transition form factors of  $\gamma^* \gamma \to f_2(1270)$, and derive the one-loop factorization formulas for the correlation function \eqref{eq:2.1} at the leading power in $1/Q^2$. By implementing ultraviolet (UV) renormalization and infrared (IR) subtraction to remove divergences, we obtain finite one-loop amplitudes and further extract the perturbative hard matching coefficients, ultimately constructing the NLL LCSRs for the hadronic photon effect.
\addtocontents{toc}{\protect\setcounter{tocdepth}{1}}  
\subsubsection{Electromagnetic current vertex diagram}
The one-loop QCD correction to the diagram $(a)$ in figure~\ref{fig:2} can be written as
\begin{align}
\Pi_{\mu\rho\sigma\alpha }^{(1,a)} =& \frac{g_{\text{em}} \, Q_q  \, g_s^2 \, C_F}{(p - k_2)^2 + i0}\, \mu^{2\epsilon} \int \frac{d^D l}{(2\pi)^D} \frac{1}{\left[(p - k_2 + l)^2 + i0\right] \left[(k_1 + l)^2 + i0\right] \left[l^2 + i0\right]} \nonumber \\[0.5em]
& \bar{v}(k_2) \, \sigma_{\rho \sigma} \, (p - 2k_2)_{\alpha} (\slashed{p} - \slashed{k}_2) \,  \gamma_\tau \,  (\slashed{p} - \slashed{k}_2 + \slashed{l}) \,  \gamma_\mu \,  (\slashed{k}_1+ \slashed{l}) \,  \gamma^\tau \,  u(k_1) \, ,
\end{align}
where $D=4-2\epsilon$. According to the power counting scheme specified in  \eqref{eq:2.8}, the leading-power expansion of the scalar integral
\begin{align}
I_1=\int \frac{d^D l}{(2\pi)^D} \frac{1}{\left[(p - k_2 + l)^2 + i0\right] \left[(k_1 + l)^2 + i0\right] \left[l^2 + i0\right]} \, ,
\end{align}
receives potential contributions from the hard and collinear regions. The hard region provides the short-distance contribution to the matching coefficient, whereas the collinear contribution becomes scaleless in dimensional regularization and is removed by the corresponding infrared subtraction in the matching procedure. After reducing the Dirac algebra of the amplitude $\Pi_{\mu\rho\sigma\alpha}^{(1,a)}$ and performing the loop-momentum integration, we obtain the hard contributions from diagram~$(a)$ to the three projected components of the partonic correlation function
\begin{align}
\Pi_{0,\mu\rho\sigma\alpha }^{(1,a),\text{h}}=&-\frac{g_{\text{em}}}{2 \sqrt{2}\, Q^{2}} \, \frac{\alpha_{s} C_{F}}{4\pi} \, \bar{n} \cdot p \, \frac{n_{\alpha}}{2}   \frac{\bar{n}_{\rho}}{2} \frac{n_{\sigma}}{2} \frac{\bar{n} \cdot p}{2} \sum_{q=u,d}Q_q \, \langle O_{\mu}(z, z^{\prime}) \rangle ^{(0)} * \bigg \{ \frac{1}{z^{\prime} \, r + \bar{z}^{\prime}} \, \frac{1}{z^{\prime} \, \bar{r}} \nonumber \\[0.5em]
& \times  \bigg ( \frac{1}{\epsilon} \left [ 4 \, z^{\prime} \, \bar{r} + 8\ln(z^{\prime} \,  r+ \bar{z}^{\prime}) \right] + z^{\prime} \, \bar{r} \, \ln\left(\frac{\mu^{2}}{Q^{2}}\right) \left[ 4 + 8 \, \ln(z^{\prime} \, r + \bar{z}^{\prime}) \right] \nonumber\\[0.5em]
&+ \left[ 16 \, z^{\prime} \, \bar{r} + 4(2 - z^{\prime} \, \bar{r})(1 - 3 \, z^{\prime} \, \bar{r}) \ln(z^{\prime} \, r + \bar{z}^{\prime}) - 4 \, z^{\prime} \, \bar{r} \, \ln^2(z^{\prime} \, r + \bar{z}^{\prime}) \right] \bigg ) \bigg \} \, , \nonumber  \\[0.5em]
\Pi_{1,\mu\rho\sigma\alpha }^{(1,a),\text{h}}=& -\frac{g_{\text{em}}}{\sqrt{2} \, Q^2} \, \frac{\alpha_s C_F}{4\pi} \, \bar{n} \cdot p \, \frac{n_\alpha}{2} \, \frac{\bar{n}_\sigma}{2} \, \frac{\bar{n}_\mu}{2} \, \frac{n \cdot (p - k_2)}{2} \sum_{q=u,d}Q_q \, \langle O_{\rho}(z, z^{\prime}) \rangle ^{(0)} * \, \bigg \{\frac{1}{z^{\prime} \, r + \bar{z}^{\prime}} \nonumber \\[0.5em]
&\times  4\left[ \frac{1}{\epsilon} + 1 + \ln\left(\frac{\mu^2}{Q^2}\right) + \frac{z^{\prime} \, r + \bar{z}^{\prime}}{z^{\prime} \,\bar{r}} \, \ln(z^{\prime} \, r + \bar{z}^{\prime}) \right] \bigg \} \, , \nonumber \\[0.5em]
g^{\mu\rho}_{\perp}\Pi_{2,\mu\rho\sigma\alpha }^{(1,a),\text{h}}=& 0,
\end{align}
 where $g^{\mu\rho}_{\perp}=g^{\mu\rho}-n^{\mu}\bar{n}^{\rho}/2-n^{\rho}\bar{n}^{\mu}/2$. For diagram~$(a)$, the vanishing of amplitude $g_\perp^{\mu\rho}\Pi_{2,\mu\rho\sigma\alpha}^{(1,a),\mathrm{h}}$ follows from the transverse projection of $(p-2k_2)_\alpha$ associated with $j_{2,\rho\sigma\alpha}$, since the external momenta $p$ and $k_2$ have no transverse components.
\subsubsection{Tensor meson vertex diagram}
Adopting the same calculation method, we can write down the one-loop amplitude for the QCD corrections to the tensor meson vertex diagram $(b)$ shown in figure~\ref{fig:2}
\begin{align}
\Pi_{\mu\rho\sigma\alpha }^{(1,b)}=&\frac{g_{\text{em}} \, Q_q \, g_s^2 \, C_F }{(p - k_2)^2 + i0} \, \mu^{2\epsilon} \int \frac{d^D l}{(2\pi)^D} \frac{1}{\left[(l + p)^2 + i0\right] \left[(l + k_2)^2 + i0\right] \left[l^2 + i0\right]} \nonumber \\[0.5em]
& \bar{v}(k_2) \, \gamma_\tau \, \slashed{l} \, \sigma_{\rho \sigma} (p + 2l)_\alpha (\slashed{p} + \slashed{l}) \, \gamma^\tau \, (\slashed{p}- \slashed{k}_2 ) \, \gamma_\mu \, u(k_1) \, .
\end{align}
We obtain the hard contributions from the one-loop diagram~$(b)$ as follows. The amplitude $\Pi_{1,\mu\rho\sigma\alpha}^{(1,b),\mathrm{h}}$ only contributes at $\mathcal{O}(\epsilon)$, vanishing in four-dimensional space:
\begin{align}
\Pi_{0,\mu\rho\sigma\alpha}^{(1 ,b),\text{h}}=& \frac{g_{\text{em}} }{2\sqrt{2} \, Q^2} \, \frac{\alpha_s \, C_F}{4\pi} \, \bar{n} \cdot p \, \frac{n_\alpha}{2} \, \frac{\bar{n}_\rho}{2} \, \frac{n_\sigma}{2} \, \frac{\bar{n} \cdot p}{2} \sum_{q=u,d}Q_q \, \langle O_{\mu}(z,z^{\prime}) \rangle ^{(0)} *  \bigg \{ \frac{1}{  z^{\prime} \, r +\bar{z}^{\prime} } \, \frac{1}{\bar{z}^{\prime} \, \bar{r}} \nonumber \\[0.5em]
& \times \bigg ( 8 \, r \, \ln\frac{z^{\prime} \, r + \bar{z}^{\prime}}{r} \left[ \frac{1}{\epsilon} + \ln\left(\frac{\mu^2}{Q^2}\right) \right] -\frac{4}{3} \bigg [ 2 \, \bar{z}^{\prime} \, \bar{r} + 6 \, r \, (\ln r - 2)\ln\frac{z^{\prime} \, r + \bar{z}^{\prime}}{r} \nonumber \\[0.5em]
&+ 3 \, r \, \ln^2\frac{z^{\prime} \, r + \bar{z}^{\prime}}{r} \bigg] \bigg ) \bigg \}\, , \nonumber \\[0.5em]
\Pi_{1,\mu\rho\sigma\alpha}^{(1,b),\text{h}}=& \, 0  \, , \nonumber \\[0.5em]
g^{\mu\rho}_{\perp}\Pi_{2,\mu\rho\sigma\alpha}^{(1,b),\text{h}}=& -\frac{g_{\text{em}}}{\sqrt{2}}  \, \frac{\alpha_s \, C_F}{4\pi} \, \frac{\bar{n}_{\sigma}}{2} \sum_{q=u,d}Q_q \, \langle O_{\alpha}(z,z^{\prime}) \rangle ^{(0)} * \left ( \frac{2}{3} \,  \frac{r- \bar{z}^{\prime} \, \bar{r}}{z^{\prime} \, r + \bar{z}^{\prime}}  \right ) \, .
\end{align}
\subsubsection{Box diagram}
The one-loop QCD correction to the diagram $(c)$ in figure~\ref{fig:2} can be written as
\begin{align}
\Pi_{\mu\rho\sigma\alpha }^{(1,c)}=& g_{\text{em}} \, Q_q \, g_s^2 \, C_F \, \mu^{2\epsilon} \int \frac{d^D l}{(2\pi)^D} \frac{1}{\left[(l + p)^2 + i0\right] \left[(l + p^{\prime})^2 + i0\right] \left[(l + k_2)^2 + i0\right] \left[l^2 + i0\right]}  \nonumber \\[0.5em]
& \bar{v}(k_2) \, \gamma_\tau \, \slashed{l} \, \sigma_{\rho \sigma} \, (p + 2l)_\alpha (\slashed{l} + \slashed{p} ) \, \gamma_\mu \, (\slashed{l}  + \slashed{p}^{\prime} ) \, \gamma^\tau \, u(k_1) \,,
\end{align}
the hard contributions from the one-loop diagram $(c)$ are as follows. Evaluating the contribution from $\Pi_{0,\mu\rho\sigma\alpha}^{(1,c),\text{h}}$ and $\Pi_{1,\mu\rho\sigma\alpha}^{(1,c),\text{h}}$, we find that the corresponding hard coefficients only contribute at $\mathcal{O}(\epsilon)$ and consequently vanish in four-dimensional space:
\begin{align}
\Pi_{0,\mu\rho\sigma\alpha}^{(1,c),\text{h}}&=0 \, ,\qquad \Pi_{1,\mu\rho\sigma\alpha }^{(1,c),\text{h}} =0 \, , \nonumber \\[0.5em]
g^{\mu\rho}_{\perp}\Pi_{2,\mu\rho\sigma\alpha }^{(1,c),\text{h}}&=-\frac{ g_{\text{em}}}{\sqrt{2}}  \, \frac{\alpha_s \, C_F}{4\pi} \, \frac{\bar n_{\sigma}}{3} \int_0^1dz^{\prime}\sum_{q=u,d}Q_q \, \langle O_{\alpha}(z,z^{\prime}) \rangle ^{(0)}  \, .
\end{align}

\subsubsection{Wave function renormalization}
The one-loop QCD correction to the diagram $(d)$ in figure~\ref{fig:2} can be written as
\begin{align}
\Pi_{\mu\rho\sigma\alpha }^{(1,d)}=& \frac{g_{\text{em}} \, Q_q \,   g_s^2 \, C_F}{\left[(p - k_2)^2 + i0\right]^2}  \,  \mu^{2\epsilon} \int \frac{d^D l}{(2\pi)^D} \frac{1}{\left[(p - k_2 + l)^2 + i0\right] \left[l^2 + i0\right]} \nonumber \\[0.5em]
& \bar{v}(k_2) \, \sigma_{\rho \sigma} \, (p - 2k_2)_\alpha (\slashed{p} - \slashed{k}_2 ) \,  \gamma_\tau \, (\slashed{p}- \slashed{k}_2 + \slashed{l}) \, \gamma^\tau \, (\slashed{p}- \slashed{k}_2) \, \gamma_\mu \, u(k_1) \,,
\end{align}
after evaluating the loop integral and extracting the hard contributions, we present the results for the one-loop diagram $(d)$ as follows. As in diagram~$(a)$, $g_\perp^{\mu\rho}\Pi_{2,\mu\rho\sigma\alpha}^{(1,d),\mathrm{h}}$ vanishes because $(p-2k_2)_\alpha^\perp=0$ in the adopted collinear kinematics:
\begin{align}
\Pi_{0,\mu\rho\sigma\alpha}^{(1,d),\text{h}}=& -\frac{\sqrt{2} \, g_{\text{em}}}{Q^2} \, \frac{\alpha_s \, C_F}{4\pi} \, \bar{n} \cdot p \, \frac{n_\alpha}{2} \, \frac{\bar{n}_\rho}{2} \, \frac{n_\sigma}{2} \, \frac{\bar{n} \cdot p}{2}  \sum_{q=u,d}Q_q \, \langle O_{\mu}(z,z^{\prime}) \rangle ^{(0)} * \bigg \{ \frac{1}{z^{\prime} \, r + \bar{z}^{\prime} }\nonumber \\[0.5em]
& \times \left[ \frac{1}{\epsilon} + 1 + \ln\left(\frac{\mu^2}{Q^2}\right) - \ln(z^{\prime} \, r + \bar{z}^{\prime}) \right] \bigg \} \, , \nonumber \\[0.5em]
\Pi_{1,\mu\rho\sigma\alpha}^{(1,d),\text{h}}=& \frac{g_{\text{em}} }{\sqrt{2} \, Q^2} \, \frac{\alpha_s \, C_F}{4\pi} \, \bar{n} \cdot p \, \frac{n_\alpha}{2} \, \frac{\bar{n}_\sigma}{2} \, \frac{\bar{n}_\mu}{2} \, \frac{n \cdot (p - k_2)}{2} \sum_{q=u,d}Q_q \, \langle O_{\rho}(z,z^{\prime}) \rangle ^{(0)} * \bigg \{ \frac{1}{z^{\prime} \, r+ \bar{z}^{\prime}} \nonumber \\[0.5em]
& 4\left[ \frac{1}{\epsilon} + 1 + \ln\left(\frac{\mu^2}{Q^2}\right) - \ln(z^{\prime} \, r + \bar{z}^{\prime}) \right] \bigg \} \, , \nonumber \\[0.5em]
g^{\mu\rho}_{\perp}\Pi_{2,\mu\rho\sigma\alpha}^{(1,d),\text{h}}=& \, 0 \,.
\end{align}

\subsubsection{Vertex emission diagrams}
In this part, we evaluate the vertex emission diagrams, labeled $(e)-(g)$ in figure~\ref{fig:2}. These diagrams originate from the gauge field term contained in the covariant derivative of the tensor-meson interpolating current $j_{\rho\sigma\alpha}$, which generates an additional quark-antiquark-gluon vertex at the insertion of the interpolating current.

The one-loop QCD correction to the diagram $(e)$ in figure~\ref{fig:2} can be written as
\begin{align}
\Pi_{\mu\rho\sigma\alpha }^{(1,e)}=&-2 \, g_{\text{em}} \,  Q_q \, g_s^2 \,  C_F \, \mu^{2\epsilon} \int \frac{d^D l}{(2\pi)^D} \frac{1}{\left[(p - k_2 + l)^2 + i0\right] \left[(k_1 + l)^2 + i0\right] \left[l^2 + i0\right]} \nonumber \\[0.5em]
& \bar{v}(k_2) \, \sigma_{\rho \sigma} \, g_{\alpha \tau} \, (\slashed{p} + \slashed{l}  - \slashed{ k}_2 ) \, \gamma_\mu \, (\slashed{ k}_1 + \slashed{l}) \, \gamma^\tau \, u(k_1)\,,
\end{align}
the hard contributions from the one-loop diagram $(e)$ are as follows, where the contributions of $\Pi_{0,\mu\rho\sigma\alpha}^{(1,e),\text{h}}$ and $\Pi_{1,\mu\rho\sigma\alpha}^{(1,e),\text{h}}$ vanish in accordance with the equation of motion,
\begin{align}
\Pi_{0,\mu\rho\sigma\alpha }^{(1,e),\text{h}}=& \, 0 \, , \qquad \Pi_{1,\mu\rho\sigma\alpha }^{(1,e),\text{h}}= 0 \, , \nonumber \\[0.5em]
g^{\mu\rho}_{\perp}\Pi_{2,\mu\rho\sigma\alpha }^{(1,e),\text{h}}=&\sqrt{2} \, g_{\text{em}} \,   \frac{\alpha_s \, C_F}{4\pi} \, \frac{\bar{n}_{\sigma}}{2} \sum_{q=u,d}Q_q \, \langle O_{\alpha}(z,z^{\prime}) \rangle ^{(0)} * \bigg \{ \bigg [ \frac{1}{\epsilon} + \ln\left( \frac{\mu^2}{Q^2} \right) + \frac{\ln(z^{\prime} \, r + \bar{z}^{\prime})}{z^{\prime} \, \bar{r}}  \nonumber \\[0.5em]
& - \ln(z^{\prime} \, r + \bar{z}^{\prime}) \bigg ] \bigg \}\, .
\end{align}

The one-loop QCD correction to the diagram $(f)$ in figure~\ref{fig:2} can be written as
\begin{align}
\Pi_{\mu\rho\sigma\alpha }^{(1,f)}=& -\frac{2 \, g_{\text{em}} \, Q_q \,  g_s^2 \, C_F}{(p - k_2)^2 + i0} \,  \mu^{2\epsilon} \int \frac{d^D l}{(2\pi)^D} \frac{1}{\left[(p - k_2 + l)^2 + i0\right] \left[l^2 + i0\right]} \nonumber \\[0.5em]
& \bar{v}(k_2) \, \sigma_{\rho \sigma} \, g_{\alpha \tau} \, (\slashed{p}+ \slashed{l} - \slashed{k}_2 ) \, \gamma^\tau \,  (\slashed{p} -\slashed{k}_2) \, \gamma_\mu \, u(k_1) \, ,
\end{align}
the hard contributions from the one-loop diagram $(f)$ are as follows
\begin{align}
\Pi_{0,\mu\rho\sigma\alpha }^{(1,f),\text{h}}=& -\frac{g_{\text{em}} }{2\sqrt{2} \, Q^2} \,  \frac{\alpha_s \, C_F}{4\pi} \, \bar{n} \cdot p \, \frac{n_\alpha}{2} \, \frac{\bar{n}_\rho}{2} \, \frac{n_\sigma}{2} \, \frac{\bar{n} \cdot p}{2} \sum_{q=u,d}Q_q \, \langle O_{\mu}(z,z^{\prime}) \rangle ^{(0)}* \bigg \{\frac{1}{z^{\prime} \, r + \bar{z}^{\prime}}  \nonumber  \\[0.5em]
&\times 8\left[ \frac{1}{\epsilon} + 2 + \ln\frac{\mu^2}{Q^2} - \ln(z^{\prime} \, r + \bar{z}^{\prime}) \right] \bigg \} \, , \nonumber \\[0.5em]
\Pi_{1,\mu\rho\sigma\alpha }^{(1,f),\text{h}}=& \, 0 \, , \nonumber \\[0.5em]
g^{\mu\rho}_{\perp}\Pi_{2,\mu\rho\sigma\alpha }^{(1,f),\text{h}}=&  \sqrt{2} \, g_{\text{em}} \, \frac{\alpha_s \, C_F}{4\pi} \, \frac{\bar{n}_{\sigma}}{2} \int_0^1dz^{\prime}\sum_{q=u,d}Q_q \, \langle O_{\alpha}(z,z^{\prime}) \rangle ^{(0)} \, .
\end{align}
 
The one-loop QCD correction to the diagram $(g)$ in figure~\ref{fig:2} can be written as
\begin{align}
\Pi_{\mu\rho\sigma\alpha }^{(1,g)}=&-\frac{2 \, g_{\text{em}} \, Q_q \,  g_s^2 \, C_F}{(p - k_2)^2 + i0} \,  \mu^{2\epsilon} \int \frac{d^D l}{(2\pi)^D} \frac{1}{\left[(l + k_2)^2 + i0\right] \left[l^2 + i0\right]} \nonumber \\[0.5em]
& \bar{v}(k_2) \, \gamma^{\tau} \, \slashed{l} \, \sigma_{\rho \sigma} \, g_{\alpha \tau} \,
(\slashed{p} - \slashed{k}_2 ) \, \gamma_\mu \, u(k_1) \, ,
\end{align}
the scalar integral of this diagram is
\begin{align}
I&=\int \frac{d^D l}{(2\pi)^D} \frac{1}{\left[(l + k_2)^2 + i0\right] \left[l^2 + i0\right]} =\int \frac{d^D l}{(2\pi)^D} \frac{1}{[l^2 + n \cdot k_2 \, \bar{n} \cdot l + i0][l^2 + i0]} \,.
\end{align}
After Feynman parametrization and a shift of the loop momentum, the scalar integral becomes scaleless in dimensional regularization and therefore vanishes. Consequently, diagram $(g)$ gives no hard contribution to any of the projected components:
\begin{align}
\Pi_{0,\mu\rho\sigma\alpha }^{(1,g),\text{h}}= 0 \, , \qquad \Pi_{1,\mu\rho\sigma\alpha }^{(1,g),\text{h}}=0 \, , \qquad g^{\mu\rho}_{\perp}\Pi_{2,\mu\rho\sigma\alpha }^{(1,g),\text{h}}=0 \, .
\end{align}
\addtocontents{toc}{\protect\setcounter{tocdepth}{2}}   
\subsubsection{The NLL LCSR for $\gamma^*\gamma \to f_2(1270)$ form factors}
We sum all the contributions to obtain the one-loop QCD corrections to the four-point amplitudes
\begin{align}
\Pi_{0,\mu\rho\sigma\alpha }^{(1)}=& \frac{\sqrt{2} \, g_{\text{em}} }{Q^{2}}\,  \bar{n} \cdot p \, \frac{n_\alpha}{2} \, \frac{\bar{n}_{\rho}}{2} \, \frac{n_{\sigma}}{2} \, \frac{\bar{n} \cdot p}{2} \, \sum_{q=u,d}  Q_{q} \,  \langle O_{\mu}(z, z') \rangle ^{(0)}  * A_{0,\text{h}}^{(1)}(z') \, , \nonumber \\[0.5em]
\Pi_{1,\mu\rho\sigma\alpha }^{(1)}=& \sqrt{2} \, g_{\text{em}}  \,  \frac{n_\alpha}{2} \, \frac{\bar{n}_{\sigma}}{2} \, \frac{\bar{n}_{\mu}}{2} \, \sum_{q=u,d}  Q_{q} \,  \langle O_{\rho}(z, z') \rangle ^{(0)}  * A_{1,\text{h}}^{(1)}(z') \, ,\nonumber \\[0.5em]
g^{\mu\rho}_{\perp}\Pi_{2,\mu\rho\sigma\alpha}^{(1)}=& -\sqrt{2} \, g_{\text{em}} \, \frac{\bar{n}_{\sigma}}{2} \, \sum_{q=u,d}  Q_{q} \,  \langle O_{\alpha}(z, z') \rangle ^{(0)}  * A_{2,\text{h}}^{(1)}(z') \, ,
\end{align}
the hard amplitudes $A_{i,\text{h}}^{(1)}(z') \, (i=0,1,2)$ are given by
\begin{align}
\label{eq:3.27}
A_{0,\text{h}}^{(1)}(z')=& \frac{\alpha_s \, C_F}{4\pi} \, \bigg\{ \frac{1}{z' \,r + \bar{z}'} \, \frac{1}{\bar{r} \, \bar{z}' \, z'} \bigg [ \frac{1}{\epsilon} \left( -4 \, \bar{r} \, \bar{z}' \, z' + 2(z' \, r - \bar{z}') \ln(z' \, r + \bar{z}') - 2 \, z ' \, r \ln r \right) \nonumber \\[0.5em]
&+ \ln\frac{\mu^2}{Q^2} \left (-4 \, \bar{r} \, \bar{z}' \, z' + 2(z' \, r - \bar{z}') \ln( z' \, r+\bar{z}' ) - 2 \, z' \, r  \ln r \right ) - \frac{29}{3} \, \bar{r} \, \bar{z}' \, z'  \nonumber \\[0.5em]
& - 3 \, \bar{z}' \, \ln(z' \, r + \bar{z}') + 4 \, z' \, r \ln(z' \, r + \bar{z}') - (z' \, r - \bar{z}') \ln^2(z' \, r + \bar{z}')  \nonumber\\[0.5em]
& + z' \, r \ln^2 r - 4 \, z' \, r \ln r + 4 \, \bar{r} \, \bar{z}' \, z' \ln(z' \, r + \bar{z}') \bigg ]  + (z' \leftrightarrow \bar{z}') \bigg \} \, , \nonumber\\[0.5em]
A_{1,\text{h}}^{(1)}(z')=&  \frac{\alpha_s \, C_F}{4\pi} \, \bigg[ \frac{1}{z' \, \bar{r}} \ln(z' \, r+ \bar{z}') + (z' \leftrightarrow \bar{z}') \bigg] \, , \nonumber \\[0.5em]
A_{2,\text{h}}^{(1)}(z')=& \frac{\alpha_s \, C_F}{4\pi} \bigg[ - \frac{1}{\epsilon} -\ln\frac{\mu^2}{Q^2} -1+\frac{2 \, r}{3(\bar{z}'+z' \, r)}-\frac{ (\bar{z}'+z' \, r) }{z' \, \bar{r}}\ln (\bar{z}'+z' \, r) + (z' \leftrightarrow \bar{z}') \bigg] \, .
\end{align}

Having completed the calculation of the one-loop hard amplitudes $A_{i,\text{h}}^{(1)}$, we now extract the perturbative hard matching coefficients $H_i^{(1)}$.  Since the amplitudes still contain UV and IR divergences, we implement UV renormalization and IR subtraction to obtain the finite, scale-dependent short-distance functions that enter the factorization formula. The matching equation for extracting the one-loop hard matching coefficient can be written as
\begin{align}
\label{eq:3.28}
 A_{i}^{(1)}(z') * \langle O_{\mu}(z, z') \rangle^{(0)} =  H_{i}^{(1)}(z') * \langle O_{\mu}(z, z') \rangle^{(0)} + H_{i}^{(0)}(z') * \langle O_{\mu}(z, z') \rangle^{(1)} \, ,
\end{align}
where $A_i^{(1)}$ is the QCD four-point one-loop amplitude in dimensional regularization, $H_i^{(1)}$ is the corresponding short-distance hard contribution, $H_i^{(0)}$ is the tree-level hard coefficient, and $\langle O_{\mu} \rangle^{(1)} $ is the one-loop SCET operator matrix element. The UV renormalized one-loop SCET matrix element $\langle O_{\mu} \rangle^{(1)} $ is given by
\begin{align}
\label{eq:3.29}
\langle O_{\mu} \rangle^{(1)} =\left[ M_{ \text{bare}}^{(1)} + Z^{(1)} \right] \langle O_{ \mu} \rangle^{(0)} \, ,
\end{align}
where $M_{ \text{bare}}^{(1)}$ denotes the bare one-loop SCET matrix element, and $Z^{(1)}$ is the one-loop expansion coefficient of the operator UV renormalization constant. When UV and IR divergences are regulated using dimensional regularization, the massless loop integrals involved in the calculation typically yield scaleless integrals, causing the bare SCET matrix element $M_{ \text{bare}}^{(1)}$ to vanish. Substituting \eqref{eq:3.29} into the matching equation \eqref{eq:3.28}, we obtain the expression for the one-loop hard matching coefficients
\begin{align}
H_i^{(1)} = A_{i}^{(1)} -  H_i^{(0)} * Z^{(1)}=A_{i,\text{h}}^{(1),\text{reg}} \, ,
\end{align}
where $A_{i,\mathrm{h}}^{(1),\mathrm{reg}}$ denotes the regularized hard contribution to the NLO QCD matrix element as presented in~\eqref{eq:3.27}.

The factorization formulas for the vacuum-to-photon correlation function can be further derived as follows
\begin{align}
\label{eq:3.33}
G_{0,\mu\rho\sigma\alpha}(p^2, Q^2) =& -\frac{\sqrt{2} \, g_{\text{em}}^{2}(Q_u^2 + Q_d^2)}{Q^2} \, \frac{\bar{n}_{\rho} }{2} \, \frac{n_{\sigma} }{2} \, \frac{n_{\alpha} }{2} \, \bar{n} \cdot p \, \frac{\bar{n} \cdot p}{2} \,  \chi(\mu) \, \langle \bar{q} q \rangle (\mu) \, n \cdot p^{\prime} \,\epsilon_{\mu}^{\perp}(p^{\prime})  \nonumber \\[0.5em]
&\int_{0}^{1} dz \left[ H_0^{(0)}(z) + H_0^{(1)}(z, \mu) \right] \phi_\gamma(z, \mu) + \mathcal{O}(\alpha_s^2) \, , \nonumber \\[0.5em]
G_{1,\mu\rho\sigma\alpha}(p^2, Q^2) =& -\sqrt{2} \, g_{\text{em}}^{2}(Q_u^2 + Q_d^2)  \, \frac{\bar{n}_{\mu} }{2} \,  \frac{\bar{n}_{\sigma} }{2} \,  \frac{n_{\alpha} }{2} \,   \chi(\mu) \, \langle \bar{q} q \rangle (\mu) \, n \cdot p^{\prime} \, \epsilon_{\rho}^{\perp}(p^{\prime})  \nonumber \\[0.5em]
&\int_{0}^{1} dz \left[ H_1^{(0)}(z) + H_1^{(1)}(z, \mu) \right] \phi_\gamma(z, \mu) + \mathcal{O}(\alpha_s^2) \, , \nonumber \\[0.5em]
g_{\perp}^{\mu\rho}G_{2,\mu\rho\sigma\alpha}(p^2, Q^2) =& \sqrt{2}\, g_{\text{em}}^{2}(Q_u^2 + Q_d^2)  \, \frac{\bar{n}_{\sigma} }{2} \, \chi(\mu) \, \langle \bar{q} q \rangle (\mu) \, n \cdot p^{\prime} \, \epsilon_{\alpha}^{\perp}(p^{\prime})  \nonumber \\[0.5em]
&\int_{0}^{1} dz \left[ H_2^{(0)}(z) + H_2^{(1)}(z, \mu) \right] \phi_\gamma(z, \mu) + \mathcal{O}(\alpha_s^2) \, .
\end{align}
We next examine the factorization-scale dependence of the
factorization formula in \eqref{eq:3.33}. For this purpose, we combine the explicit scale dependence of the hard matching coefficients with the renormalization-group (RG) evolution of the leading twist photon matrix element, whose evolution equation reads
\begin{align}
\mu^2 \frac{d}{d\mu^2} \left[ \chi(\mu) \, \langle \bar{q} q \rangle (\mu) \, \phi_\gamma(z, \mu) \right]  = \int_0^1 dz' \, V(z, z') \, \left[ \chi(\mu) \, \langle \bar{q} q \rangle (\mu) \, \phi_\gamma(z', \mu) \right] \, ,
\end{align}
where the perturbative expansion form of the evolution kernel $V(z,z')$ is
\begin{align}
V(z, z') =\sum_{n=0}^{\infty} \left( \frac{\alpha_s}{4\pi} \right)^{n+1} V_n(z, z') \,.
\end{align}
The one-loop renormalization kernel $V_0(z, z')$ is given by~\cite{Lepage:1979zb, Shifman:1980dk}
\begin{align}
V_0(z, z') = 2 \, C_F \left[ \frac{\bar{z}}{\bar{z'}} \, \frac{1}{z - z'} \, \theta(z - z') + \frac{z}{z'} \, \frac{1}{z' - z} \, \theta(z' - z) \right]_+ - C_F \, \delta(z - z') \, ,
\end{align}
where the plus function is defined as
\begin{align}
\left[ f(z,z') \right]_+ = f(z,z')-\delta(z-z') \int_{0}^{1} dt \, f(t,z') \, .
\end{align}
Combining these results, we find that
\begin{align}
\frac{d}{d\ln \mu} \Pi _{0,\mu\rho\sigma\alpha} =& -\frac{1}{2} \, \frac{\alpha_{s} \, C_{F}}{\pi}  \, \Pi _{0,\mu\rho\sigma\alpha}^{\,(0)} + \mathcal{O}(\alpha_s^2) \, , \nonumber \\[0.5em]
\frac{d}{d\ln \mu} \Pi _{1,\mu\rho\sigma\alpha} =& -\frac{1}{2} \, \frac{\alpha_{s} \, C_{F}}{\pi} \, \Pi _{1,\mu\rho\sigma\alpha}^{\,(0)} + \mathcal{O}(\alpha_s^2) \, , \nonumber \\[0.5em]
\frac{d}{d\ln \mu} g_{\perp}^{\mu\rho}\Pi _{2,\mu\rho\sigma\alpha} =& 4\,  g_{\mathrm{em}}^2 \, 
(Q_u^2 + Q_d^2) \, \frac{\bar{n}_{\sigma}}{2} \, \int_0^1 dz \, (-2  \,  \frac{\alpha_{s} \, C_{F}}{4 \, \pi} ) \,  \chi(\mu) \, \langle \bar{q} q \rangle (\mu) \nonumber \\[0.5em]
& \times n \cdot p^{\prime} \, \epsilon_\alpha^\perp \, \phi_\gamma(z, \mu) \, + \mathcal{O}(\alpha_s^2) \, .
\end{align}
The evolution equation for the renormalization constant of the interpolating current $j_{\rho\sigma\alpha}$ given in~\cite{Cheng:2010hn}, reads
\begin{align}
\frac{d}{d\ln\mu} \ln f_{f_2}^{\perp}(\mu) = -2 \, \sum_{n=0}^{\infty} \left( \frac{\alpha_s(\mu)}{4\pi} \right)^{n+1} \gamma^{\perp(n)}, \quad \gamma^{\perp(0)} = 3 \, C_F \,.
\end{align}
The residual factorization-scale dependence originates from the renormalization of the tensor interpolating current adopted in this work. The current $j_{\rho\sigma\alpha}$ can mix with other operators carrying the same quantum numbers. A complete cancellation of the factorization-scale dependence at the level of the correlation function would require the anomalous-dimension matrix $\gamma_{ij}$ for the full operator basis. In the present analysis, we include only the diagonal anomalous dimension associated with the tensor interpolating current.

Next, we perform the NLL resummation for the large logarithmic terms $\mathcal{O}(\ln(\mu^2/Q^2))$ in $A_{i,\text{h}}^{(1)}$. The two-loop coefficient of the evolution kernel $V(z,z')$ is given by~\cite{Belitsky:1999gu,Mikhailov:2008my,Belitsky:2000yn}, and the  explicit expressions of the kernel functions can be found in~\cite{Mikhailov:2008my}
\begin{align}
V_1(z,z') = \frac{N_f}{2} C_F V_N(z,z') + C_F C_A V_G(z,z') + C_F^2 V_F(z,z') \,,
\end{align}
the Gegenbauer polynomial series expansion form of the leading-twist
photon DA $\phi_{\gamma}(z,\mu)$~\cite{Ball_2003} is
\begin{align}
\phi_{\gamma}(z,\mu)=6 \, z \, \bar{z} \sum_{n=0}^{\infty} a_{n}(\mu) \, C_{n}^{3/2}(2z-1) \, ,
\end{align}
where $C_n^{(a)}(z)$ denotes the Gegenbauer polynomial of order $n$. The two-loop evolution 
of the Gegenbauer moment $ a_n(\mu)$  can be obtained as follows
\begin{align}
\chi(\mu)\,\langle\bar{q}q\rangle(\mu)\,a_n(\mu) =& E_{T,n}^{\mathrm{NLO}}(\mu,\mu_0)\,\chi(\mu_0)\,\langle\bar{q}q\rangle(\mu_0)\,a_n(\mu_0) \nonumber\\[0.5em]
&+\frac{\alpha_s(\mu)}{4\pi}\sum_{k=0}^{n-2} E_{T,n}^{\mathrm{LO}}(\mu,\mu_0)\,d_{T,n}^{k}(\mu,\mu_0)\,\chi(\mu_0)\,\langle\bar{q}q\rangle(\mu_0)\,a_k(\mu_0) \,,
\end{align}
where $k,n=0,2,4,\ldots	$,  and the explicit expressions of $E_{T,n}^{(N)LO}$
and $d_{T,n}^{k} $ can be found  in Appendix A of \cite{Wang_2017}. Combining everything together, we obtain the following NLL resummation 
improved factorization formula for the three projected components of the correlation function:
\begin{align}
G_{0,\mu\rho\sigma\alpha}(p^2, Q^2) =& -\frac{\sqrt{2} \, g_{\text{em}}^{2}(Q_u^2 + Q_d^2)}{Q^2} \, \frac{\bar{n}_{\rho} }{2} \, \frac{n_{\sigma} }{2} \, \frac{n_{\alpha} }{2} \, \bar{n} \cdot p \, \frac{\bar{n} \cdot p}{2} \,  \chi(\mu) \, \langle \bar{q} q \rangle (\mu) \, n \cdot p^{\prime} \,\epsilon_{\mu}^{\perp}(p^{\prime})  \nonumber \\[0.5em]
&\sum_{n=0} a_n(\mu) \, C_{n,0}(p^2,Q^2) + \mathcal{O}(\alpha_s^2) \, , \nonumber \\[0.5em]
G_{1,\mu\rho\sigma\alpha}(p^2, Q^2) =& -\sqrt{2} \, g_{\text{em}}^{2}(Q_u^2 + Q_d^2)  \,  \frac{\bar{n}_{\mu} }{2} \,  \frac{\bar{n}_{\sigma} }{2} \,  \frac{n_{\alpha} }{2} \, \chi(\mu) \, \langle \bar{q} q \rangle (\mu) \, n \cdot p^{\prime} \, \epsilon_{\rho}^{\perp}(p^{\prime}) \nonumber  \\[0.5em]
&\sum_{n=0} a_n(\mu) \, C_{n,1}(p^2,Q^2) + \mathcal{O}(\alpha_s^2) \, , \nonumber\\[0.5em]
g_{\perp}^{\mu\rho}G_{2,\mu\rho\sigma\alpha}(p^2, Q^2) =& \sqrt{2}\, g_{\text{em}}^{2}(Q_u^2 + Q_d^2)  \, \frac{\bar{n}_\sigma}{2}  \chi(\mu) \, \langle \bar{q} q \rangle (\mu) \, n \cdot p^{\prime} \, \epsilon_{\alpha}^{\perp}(p^{\prime})  \nonumber \\[0.5em]
&\sum_{n=0} a_n(\mu) \, C_{n,2}(p^2,Q^2) + \mathcal{O}(\alpha_s^2) \, ,
\end{align}
where the explicit expression of the perturbative matching coefficient $C_{n,i}(p^2,Q^2)$ $(i=0,1,2)$ is 
\begin{align}
C_{n,i}(p^2,Q^2)=\int_{0}^{1} dz \left[ H_i^{(0)}(z) + H_i^{(1)}(z, \mu) \right] \left [ 6 \, z \, \bar{z} \, C_n^{3/2}(2z-1) \right ] \,.
\end{align}

Employing the spectral representations of the convolution integrals presented in Appendix \ref{app:A}, we derive the dispersive form of the NLL factorization formulas. By matching these expressions with the hadronic representation of the correlation function $G_{\mu\rho\sigma\alpha}(p^{\prime},q)$ in \eqref{eq:3.8} and applying the quark-hadron duality approximation, we obtain the NLL LCSRs for the hadronic photon corrections to the $\gamma^*\gamma \to f_2(1270)$ form factors at leading twist. The explicit expressions read as follows:
\begin{align}
 T_0^{\text{NLP}}(Q^2) =& - \frac{3 \sqrt{2} \, g_{\text{em}}^{2} \, (Q_u^2 + Q_d^2 ) \, e^{m_{f_2}^2 / M^2}}{4 \, f_{f_2}^{\perp} \, m_{f_2}} \,  \chi(\mu) \, \langle \bar{q} q \rangle (\mu) \int_{0}^{s_0} ds \, e^{-s / M^2} \nonumber \\[0.5em]
& \times \left [ \rho_{0}^{(0)}(s, Q^2) + \frac{\alpha_s \, C_F}{4\pi} \rho_{0}^{(1)}(s, Q^2) \right ] \frac{s+Q^2}{Q^2} \, , \nonumber\\[0.5em]
T_1^{\text{NLP}}(Q^2) =& 4\sqrt{2}\,  \frac{Q^2 \, g_{\text{em}}^2 \, ( Q_u^2 + Q_d^2) \, e^{m_{f_{2}}^{2} / M^{2}}}{f_{f_2}^{\perp} \, m_{f_{2}}^{3}}  \,  \chi(\mu) \, \langle \bar{q} q \rangle (\mu) \int_{0}^{s_0} ds \, e^{-s / M^2} \nonumber\\[0.5em]
& \times \left [ \rho_{1}^{(0)}(s, Q^2) + \frac{\alpha_s \, C_F}{4\pi} \rho_{1}^{(1)}(s, Q^2) \right ]\frac{s+Q^2}{Q^2} \, , \nonumber \\[0.5em]
T_2^{\text{NLP}}(Q^2) =& 2\sqrt{2} \,  \frac{Q^2 \, g_{\text{em}}^2 \, (Q_u^2 + Q_d^2) \, e^{m_{f_2}^2 / M^2}}{f_{f_2}^\perp \, m_{f_2}^3} \, \chi(\mu) \, \langle \bar{q} q \rangle (\mu) \int_{0}^{s_0} ds \, e^{-s / M^2} \nonumber \\[0.5em]
& \times \frac{\alpha_s \, C_F}{4\pi}\rho_{2}^{(1)}(s, Q^2) \frac{s+Q^2}{Q^2} \, ,
\end{align}
the QCD spectral densities $\rho_{i}^{(j)}(s,Q^2)(i=0,1,2$ and $j=0,1)$ can be written as
\begin{align}
\rho_{0}^{(0)}(s, Q^2) =&\frac{Q^2}{Q^2 + s}  \, \phi_\gamma\left( \frac{Q^2}{Q^2 + s}, \mu \right) \, , \nonumber\\[0.5em]
\rho_{0}^{(1)}(s, Q^2) =& 2 \int_{0}^{1} \frac{dz}{\bar{z}} \bigg\{ \theta\left(z - \frac{Q^2}{Q^2 + s}\right) \frac{Q^2}{Q^2 + s} \left[ \frac{\bar{z} - z}{z} \ln\left( \frac{\mu^2}{z \, s - \bar{z} \,  Q^2}\right) + \frac{3 \, \bar{z}}{2 \, z} - 2 \right]  \nonumber \\[0.5em]
& + \left[ \ln\left( \frac{\mu^2}{s} \right) + 2 \right] \left[ \frac{Q^2}{s + Q^2} - \mathcal{P} \frac{\bar{z} \,  Q^2}{\bar{z} \, Q^2 - z \, s} \right]
\bigg\} \, \phi_\gamma(z, \mu) \nonumber \\[0.5em]
& + \frac{Q^2}{Q^2 + s} \int_{0}^{1} dz \, \, \theta\left(z - \frac{Q^2}{Q^2 + s}\right) \bigg\{ 2 \, \ln\left ( \frac{z \, s - \bar{z} \, Q^2}{Q^2} \right )  \bigg [  \ln\left (  \frac{\mu^2}{z \, s - \bar{z} \, Q^2} \right ) \nonumber  \\[0.5em]
&  +\ln\left (  \frac{\mu^2}{Q^2} \right )  + \frac{3}{2}  \bigg ] - \ln^2\left ( \frac{\mu^2}{Q^2} \right )  + \ln^2\left ( \frac{\mu^2}{s}  \right ) - 4 \, \ln\left ( \frac{s}{\mu^2}  \right ) \nonumber \\[0.5em]
& - \frac{\pi^2}{3} + \frac{29}{3}
\bigg \} \, \frac{d}{dz} \phi_\gamma(z, \mu) \, , \nonumber \\[0.5em]
\rho_{1}^{(0)}(s, Q^2) =& 0 \, , \nonumber \\[0.5em]
\rho_{1}^{(1)}(s, Q^2) =& - \int_{0}^{1}  \, \frac{dz}{z} \, \theta\left(z - \frac{Q^2}{s + Q^2}\right)  \,  \frac{Q^2}{Q^2 + s}  \, \phi_\gamma(z, \mu) \, , \nonumber \\[0.5em]
\rho_{2}^{(0)}(s, Q^2) =& 0 \, , \nonumber \\[0.5em]
\rho_{2}^{(1)}(s, Q^2) =& -\frac{2}{3} \, \frac{s}{Q^2 + s} \, \phi_\gamma\left( \frac{Q^2}{s + Q^2}, \mu \right) \nonumber \\[0.5em]
&  +  \int_{0}^{1} dz \left( \frac{1}{z} \, \frac{Q^2}{Q^2 + s} - 1 \right) \, \theta \left(z - \frac{Q^2}{s + Q^2}\right) \, \phi_\gamma(z, \mu) \, ,
\end{align}
where $\mathcal{P}$ indicates the principal-value prescription. Adopting the same power-counting scheme as in \eqref{eq:3.10}, we obtain the scaling behavior of the leading-twist hadronic-photon contributions at large $Q^2$
\begin{align}
\frac{T_{1}^{\text{NLP}}(Q^2)}{T_{1}^{\text{LP}}(Q^2)} \sim \mathcal{O}\left(\frac{\Lambda^2}{Q^2}\right) \,, \qquad \frac{T_{2}^{\text{NLP}}(Q^2)}{T_{2}^{\text{LP}}(Q^2)} \sim \mathcal{O}\left(\frac{\Lambda^2}{Q^2}\right) \,. 
\end{align}

\section{Numerical analysis}
\label{Numerical analysis}
In this section, we perform a phenomenological investigation into the impact of the hadronic photon corrections on the transition form factors for the $\gamma^* \gamma \to f_2(1270)$ process.
The main goal of this analysis is to quantify the numerical significance of the hadronic photon corrections relative to the LP contributions from QCD collinear factorization, and to present updated predictions with theoretical uncertainties for the $\gamma^* \gamma \to f_2(1270)$ form factors that can be directly confronted with experimental data. We collect several necessary input parameters, including the decay constant of the $f_2 (1270)$ meson, the photon distribution amplitude, the magnetic susceptibility $\chi(\mu)$, the quark condensate density $\langle \bar{q} q \rangle (\mu)$, and the value of the Borel parameter. Among these, the LP contributions to the three helicity amplitudes for the $\gamma^* \gamma \to f_2(1270)$ transition are adopted from the results of QCD collinear factorization presented in \cite{Braun:2016tsk}.

\subsection{Theory inputs}
\label{Theory inputs}
For comparison with the experimental measurements and previous theoretical analyses, we present the three transition form factors normalized to $T_2 (0)$. The normalization factor can be determined from the two-photon decay width of the $f_2(1270)$ meson, which is given by~\cite {ParticleDataGroup:2014cgo}
\begin{align}
\Gamma[f_2 \to \gamma\gamma] = \frac{\pi \, \alpha^2}{5 \, m_{f_2}} \left( \frac{2}{3} |T_0(0)|^2 + |T_2(0)|^2 \right) = 3.03(40)\ \text{keV} \, ,
\end{align}
where $\alpha \approx 1/137$ is the electromagnetic coupling constant. Assuming $|T_2(0)| \gg |T_0(0)|$ at $Q^2=0$, the normalization factor is estimated as 
\begin{align}
|T_2(0)| \simeq \sqrt{\frac{5  \, m_{f_2}}{\pi \, \alpha^2} \,\Gamma[f_2 \to \gamma\gamma]} = 339(22)\ \text{MeV} \, .
\end{align}
Accordingly, all our results are normalized by $T_2(0)=339 $ MeV for direct comparison with experimental data and theoretical predictions in previous works. For the numerical analysis, we use
\begin{align}
m_{f_2}=1.270~\mathrm{GeV}, \qquad f_{f_2}^{\perp}(1~\mathrm{GeV})=117(25)~\mathrm{MeV}.
\end{align}
The LP contributions adopted from
Ref.~\cite{Braun:2016tsk} depend on the leading-twist DAs of the $f_2$ meson. We therefore briefly specify
the conventions and nonperturbative inputs employed for these
distributions. Their explicit Gegenbauer parametrization reads
\begin{align}
\phi^{\parallel,\perp}(u,\mu)
=
6u(1-u)
\sum_{n=1,3,5,\ldots}
b_{n}^{\parallel,\perp}(\mu)\,
C_{n}^{3/2}(2u-1),
\end{align}
with the lowest nonvanishing Gegenbauer moments taken to be
$b_1^{\parallel}(1\,\text{GeV})=b_1^{\perp}(1\,\text{GeV})=5/3$. The scale dependence of the transverse decay constant is given by~\cite{Cheng:2010hn}
\begin{align}
f_{f_2}^{\perp} (\mu) &= f_{f_2}^{\perp}(\mu_0) \left( \frac{\alpha_s(\mu_0)}{\alpha_s(\mu)} \right)^{-\gamma_{n}^{\perp}/\beta_0} \, ,
\end{align}
where $\beta_0=(11N_c-2n_f)/3$ and the one-loop anomalous dimensions are
\begin{align}
\gamma_{n}^{\perp} = C_F \left( 1 + 4 \sum_{j=2}^{n+1} \frac{1}{j} \right) \,,
\end{align}
with $C_F=(N_c^2-1)/(2N_c)$. Following Ref.~\cite{Cheng:2010hn}, we adopt the Borel mass and the threshold parameter
\begin{align}
M^2\in[1.0, 1.4] \, \mathrm{GeV^2} \, , \qquad s_0=2.53 \, \mathrm{GeV^2} \,.
\end{align}
We adopt the value of the quark condensate density $\langle \bar{q} q \rangle (2\,\text{GeV}) = -\left( 272\pm5\,\text{MeV} \right)^3$ obtained in~\cite{FlavourLatticeAveragingGroupFLAG:2021npn}. For the magnetic susceptibility of the quark condensate, we use the QCD sum rule estimate at the reference scale $\mu_0=1\,\mathrm{GeV}$, $\chi(1\,\text{GeV}) = (3.15\pm0.3)\,\text{GeV}^{-2}$ as given in~\cite{Ball:2002ps}. Regarding the leading-twist photon DA, we adopt the estimated value $a_2(1  \, \text{GeV})=0.07\pm0.07$ for the second Gegenbauer coefficient in the numerical calculations, as determined from QCD sum rules~\cite{Ball:2006wn}.

For the LP contributions from QCD factorization, the expressions for the form factors are given by~\cite{Braun:2016tsk}. The numerical values and scale dependences of the relevant input parameters are provided in Appendix \ref{app:B}. In the numerical analysis, we adopt the values $f_q(1 \, \text{GeV})=85(10) \,\text{MeV}$,  $f_g^S(1 \, \text{GeV})=45 \,\text{MeV}$, and $f_g^T(1 \, \text{GeV})\approx 20 \,\text{MeV}$ for the quark and gluon couplings. These values are motivated by theoretical estimates from QCD sum rules and the agreement between theoretical predictions and experimental data, while remaining within reasonable theoretical uncertainties~\cite{Cheng:2010hn, Aliev:1981ju, Aliev:1982ab, Hoferichter:2020lap}. To account for possible nonasymptotic effects in the leading-twist DAs of the $f_2(1270)$, we include the contribution of the first higher Gegenbauer moment in our phenomenological analysis. Since no direct nonperturbative determination of the higher Gegenbauer moments of the $f_2(1270)$ DAs is currently available, we use the scalar-meson results of Ref.~\cite{Cheng:2005nb} only as a phenomenological guide to estimate the relative size of the first nonasymptotic conformal correction. At the reference scale $\mu_0=1\,\mathrm{GeV}$, we adopt the representative estimate $b_3 / b_1 =-0.16$, motivated by the similar ratios obtained for the light scalar states in the reference. This prescription is intended solely to model a possible departure of the tensor-meson LCDA from its asymptotic form and does not imply that the scalar and tensor-meson Gegenbauer moments are dynamically equivalent.

\begin{figure}[htbp]
    \centering

    \includegraphics[width=0.6\textwidth]{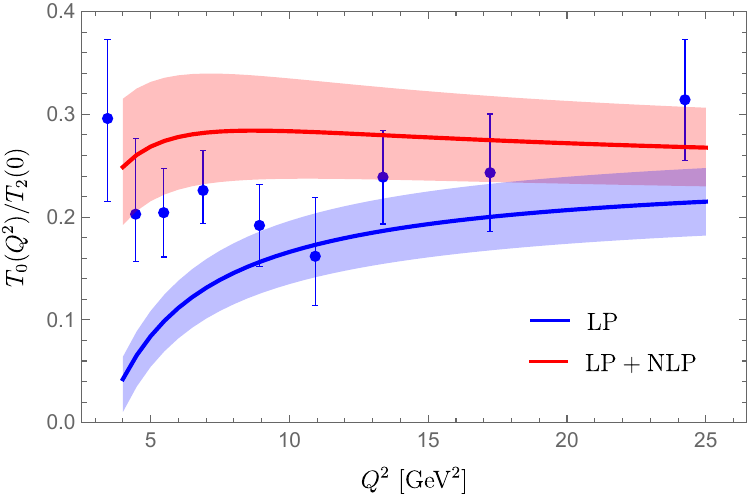}
    \par\vspace{1em}

    \includegraphics[width=0.6\textwidth]{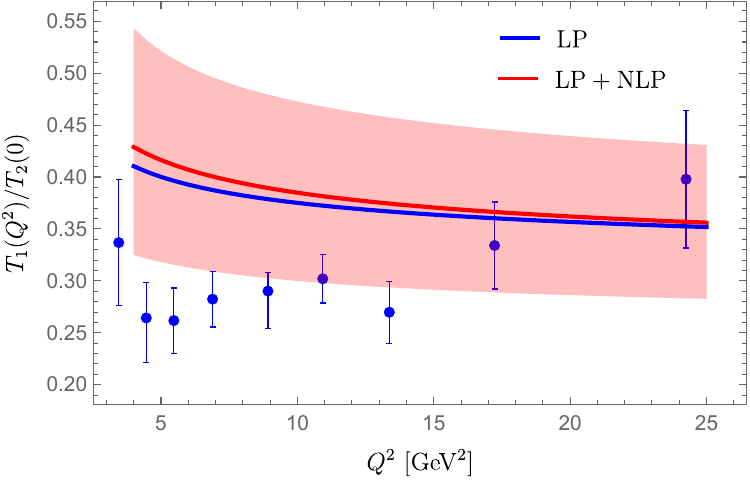}
    \par\vspace{1em}

    \includegraphics[width=0.6\textwidth]{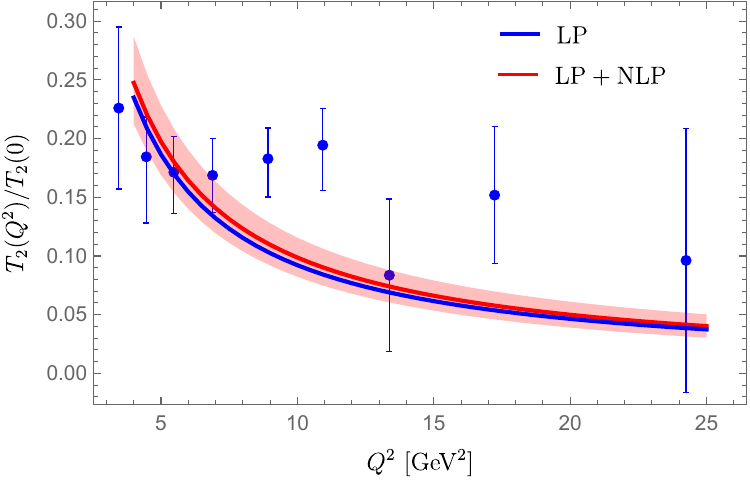}
    
    \caption{The $Q^2$ dependence of the $\gamma^* \gamma \to f_2(1270)$ form factors. The shaded area represents the combined theoretical uncertainty obtained by propagating the errors of individual parameters. The experimental data are taken from \cite{Belle:2015oin}. 
}
    \label{fig:3}
\end{figure}

\subsection{ Predictions for the $\gamma^*\gamma \to f_2(1270)$ form factors}
\label{Predictions}
This section focuses on the phenomenological impact of hadronic photon corrections to the transition form factors for $\gamma^*\gamma \to f_2(1270)$. The numerical results are shown in figure~\ref{fig:3}, with the LP contributions evaluated using the input parameters detailed in Appendix~\ref{app:B}. The LP contributions within the QCD factorization framework gradually flatten with the increase of $Q^2$, reflecting the typical asymptotic behavior of the LP term at large momentum transfer. The hadronic photon corrections provide next-to-leading-power contributions to all three form factors $T_0(Q^2)$, $T_1(Q^2)$, and $T_2(Q^2)$. In particular, the NLP correction significantly enhances the LP prediction for $T_0(Q^2)$, leading to better agreement with the experimental data. For $T_1(Q^2)$ and $T_2(Q^2)$, the corresponding hadronic photon contributions start from the $\mathcal{O}(\alpha_s)$ hard-scattering kernels, and their numerical impact remains relatively modest throughout the considered $Q^2$ region. The hadronic photon corrections to $T_0(Q^2)$, $T_1(Q^2)$, and $T_2(Q^2)$ all exhibit a power-suppressed behavior with increasing $Q^2$, consistent with the expectations of QCD factorization.

Combining the hadronic photon corrections derived above with the leading-power results from QCD factorization, we obtain the updated theoretical predictions for the transition form factors shown in figure~\ref{fig:3}:
\begin{itemize}
    \item $T_0(Q^2)/T_2(0)$: With the parameter set adopted in this work, the LP prediction from QCD factorization lies below the experimental data across the whole $Q^2$ range. After including the NLP correction from the hadronic photon effect, the theoretical prediction is shifted upward and shows better agreement with the experimental data over the entire $Q^2$ range considered. At larger values of $Q^2$, the NLP correction follows the expected $1/Q^2$ power-suppressed scaling and the theoretical prediction gradually converges toward the LP curve. This overall behavior is consistent with the theoretical expectation of power suppression in QCD factorization. Furthermore, we observe that the theoretical result exhibits a significant sensitivity to nonperturbative input parameters, such as the quark and gluon coupling parameters $f_q$ and $f_g^S$.  Allowing for variations of the parameters within their reasonable ranges can adjust the discrepancy between the LP prediction for $T_0(Q^2)$ and the experimental data, leading to improved consistency between the prediction including hadronic photon corrections and the measurements. This feature reflects the sensitivity of $T_0(Q^2)$ to non-perturbative inputs, making a more precise determination from future experiments or lattice QCD calculations highly desirable.
    \item $T_1(Q^2)/T_2(0)$: The LP prediction for $T_1(Q^2)$ from QCD factorization receives contributions from both the Wandzura--Wilczek-type terms and the genuine twist-three quark-antiquark-gluon distribution amplitudes~\cite{Braun:2016tsk}. It varies only mildly, remaining in the range 0.35–0.42 over the kinematic interval $4<Q^2<25 \, \text{GeV}^2$. The agreement between the theoretical prediction and the experimental data improves with increasing $Q^2$. With the inclusion of the hadronic photon correction obtained within the LCSR framework, the overall result shifts upward relative to the LP contribution by approximately (1–5)\% and gradually coincides with the LP contribution curve for $Q^2>10\ \text{GeV}^2$.
    \item $T_2(Q^2)/T_2(0)$: The LP prediction for $T_2(Q^2)$ decreases mildly with increasing $Q^2$, while the form factor remains dominated by the Wandzura–Wilczek-type higher-twist contribution in the phenomenologically relevant kinematic region~\cite{Braun:2016tsk}. After including the subleading power correction from the hadronic photon effect obtained within the LCSR framework, the prediction for $T_2 (Q^2)/T_2 (0)$ remains close to the LP result, with a relative shift of approximately (5–8)\%.

\end{itemize}

\section{Conclusion}
\label{Conclusion}
In this work, we calculated the NLO QCD corrections to the hadronic photon contributions to the $\gamma^{*}\gamma \to f_{2}(1270)$ transition form factors within the LCSR framework. The resummation of the large logarithms that appear in the hard matching coefficients was  performed by solving the two-loop QCD evolution equation for the light-ray tensor operator to NLL accuracy. Combining the hadronic photon corrections obtained in this work with the known LP results from QCD factorization, we provided updated theoretical predictions for the three helicity form factors. The theoretical predictions show that the form factor $T_{0}(Q^{2})$ receives substantial corrections relative to the LP contribution in the lower-$Q^2$ region. As $Q^2$ increases, we find that, for $Q^2>10\;\mathrm{GeV}^2$, the NLP contribution is less than $50\%$ of the LP result, while the relative correction decreases to approximately $20\%$ as $Q^2$ approaches $25\;\mathrm{GeV}^2$. The inclusion of the hadronic  photon contribution brings the theoretical prediction closer to the trend of the available experimental measurements and improves the overall description of the data in the considered kinematic range. For $T_1(Q^2)$ and $T_2(Q^2)$, the leading nonzero hadronic photon effects arise from the $\mathcal{O}(\alpha_s)$ hard-scattering kernels. Their numerical impact is relatively modest over the considered kinematic region. In particular, the correction to $T_1(Q^2)$ is about (1-5)\% for $4<Q^2<25~\mathrm{GeV}^2$, while the corresponding effect for $T_2(Q^2)$ is approximately (5-8)\%. 

The present study may be extended to investigate hadronic photon effects in other two-photon meson production processes. The comparison with the available experimental measurements suggests that the LCSR approach provides a reasonable description of the $\gamma^{*}\gamma \to f_2(1270)$ transition form factors in the considered kinematic region. The combined predictions are broadly compatible with the available measurements once the theoretical uncertainties are taken into account, but the comparison also reveals a strong sensitivity to several poorly constrained nonperturbative inputs. In particular, the uncertainties associated with the tensor-meson quark and gluon couplings, the higher Gegenbauer moments of the \(f_2(1270)\) DAs, the magnetic susceptibility of the quark condensate, and the leading-twist photon DA limit the precision of the present analysis. More reliable determinations of these quantities from lattice QCD, QCD sum rules, or future experimental measurements would therefore be essential for improving the phenomenological predictions. Further theoretical developments should also include higher-twist photon contributions and a more complete treatment of the renormalization and operator-mixing effects associated with the tensor-meson interpolating current.

\acknowledgments

We would like to thank Yong-Kang Huang, Bo-Xuan Shi, and Xue-Chen Zhao for helpful and illuminating discussions. This work acknowledges support from the National Natural Science Foundation of China with Grants No. 12475097 and No. 12535006, from the Natural Science Foundation of Tianjin with Grant No. 25JCZDJC01190, and from the Fundamental Research Funds for the Central Universities with Grant No. 63261180.

\appendix
\section{Spectral representations}
\label{app:A}
We present the formulas for the dispersion representations of the convolution integrals required for the calculation of correlation functions. These formulas are used to analytically continue the QCD correlation function in the variable $p^2$ to the physical region $p^2=s>0$, from which the spectral density entering the sum rule is obtained.

\begin{flalign}
\label{A.1}
&\frac{1}{\pi} \, \text{Im}_s \int_0^1 dz \, \frac{1}{z\,r+\bar{z}} \, \frac{z\, r-\bar{z}}{z \, \bar{z} \, \bar{r}} \,  \ln(z \, r+\bar{z}) \, \phi_{\gamma}(z,\mu) &   \nonumber \\
&= \frac{Q^2}{Q^2+s} \int_0^1 dz \, \theta\left(z - \frac{Q^2}{Q^2+s}\right) \left [ \frac{\bar{z}-z}{z \, \bar{z}} + 2 \, \ln \left(\frac{z \, s-\bar{z} \, Q^2}{Q^2} \right) \,  \frac{d}{dz}\right ] \phi_{\gamma}(z,\mu) \,. 
\end{flalign}
\begin{flalign}
&\frac{1}{\pi} \, \text{Im}_s \int_0^1 dz \, \frac{1}{z\,r+\bar{z}} \, \frac{r}{\bar{z} \, \bar{r}} \,  \ln r \, \phi_{\gamma}(z,\mu) & \nonumber \\
&= -\frac{Q^2}{Q^2 + s} \ln\left( \frac{s}{Q^2} \right) \,  \phi_\gamma \, \left( \frac{Q^2}{Q^2 + s}, \mu \right) - \int_0^1 \frac{dz}{\bar{z}} \left [ \frac{Q^2}{s + Q^2} - \mathcal{P} \frac{\bar{z} \, Q^2}{z \, s - \bar{z} \,  Q^2} \right ] \, \phi_\gamma(z, \mu)\,. &
\end{flalign}
\begin{flalign}
&\frac{1}{\pi} \, \text{Im}_s \int_0^1 dz \, \frac{1}{z\,r+\bar{z}} \, \frac{ 1}{z \, \bar{r}}   \, \ln (z \, r+ \bar{z}) \, \phi_{\gamma}(z,\mu) &\nonumber \\
&=-\frac{Q^2}{Q^2 + s} \int_0^1 dz \, \theta \left(z-\frac{Q^2}{Q^2 + s} \right) \left [ \frac{1}{z}+\ln \left (\frac{z \, s-\bar{z} \, Q^2}{Q^2} \right ) \, \frac{d}{dz} \right ] \, \phi_{\gamma}(z,\mu) \,. 
\end{flalign}
\begin{flalign}
&\frac{1}{\pi} \, \text{Im}_s \int_0^1 dz \, \frac{1}{z\,r+\bar{z}} \, \frac{ r}{\bar{z} \, \bar{r}} \,  \ln (z \, r+ \bar{z}) \, \phi_{\gamma}(z,\mu) &\nonumber \\
&=-\frac{Q^2}{Q^2 + s} \int_0^1 dz \, \theta \left (z-\frac{Q^2}{Q^2 + s} \right ) \left [ \frac{1}{\bar{z}}-\ln  \left (\frac{z \, s-\bar{z} \, Q^2}{Q^2} \right ) \, \frac{d}{dz} \right ] \, \phi_{\gamma}(z,\mu) \,.
\end{flalign}
\begin{flalign}
&\frac{1}{\pi} \, \text{Im}_s \int_0^1 dz \, \frac{1}{z\,r+\bar{z}} \, \frac{z \, r-\bar{z}}{z \, \bar{r} \, \bar{z}}  \, \ln^2 (z \, r+ \bar{z}) \, \phi_{\gamma}(z,\mu) &\nonumber \\
&=\frac{Q^2}{Q^2 + s} \int_0^1 dz \, \theta \left (z-\frac{Q^2}{Q^2 + s} \right ) \, \bigg \{2 \, \frac{\bar{z}-z}{z \, \bar{z}} \ln \left (\frac{z \,s- \bar{z} \, Q^2}{Q^2} \right )+ \bigg [ 2 \, \ln^2 \left (\frac{z \,s- \bar{z} \, Q^2}{Q^2} \right )  &\nonumber\\
& \quad -\frac{2}{3} \pi^2 \bigg ] \, \frac{d}{dz}\bigg \} \, \phi_{\gamma}(z,\mu) \,.& 
\end{flalign}
\begin{flalign}
&\frac{1}{\pi} \, \text{Im}_s \int_0^1 dz \, \frac{1}{z\,r+\bar{z}} \, \frac{r}{ \bar{z} \, \bar{r}} \,   \ln^2 r \, \phi_{\gamma}(z,\mu) &\nonumber \\
&= -\frac{Q^2}{Q^2 + s} \left [ \ln^2\left( \frac{s}{Q^2} \right) -\pi^2 \right ] \,\phi_\gamma\left( \frac{Q^2}{Q^2 + s}, \mu \right) - 2 \, \ln\left( \frac{s}{Q^2} \right) \, \int_0^1 \frac{dz}{\bar{z}} \bigg [ \frac{Q^2}{s + Q^2} &\nonumber \\
&- \mathcal{P} \frac{\bar{z} \, Q^2}{z \, s - \bar{z} \,  Q^2} \bigg ] \, \phi_\gamma(z, \mu) \,.
\end{flalign}
\begin{flalign}
&\frac{1}{\pi} \, \text{Im}_s \int_0^1 dz \, \frac{1}{z\,r+\bar{z}} \,  \ln(z \, r + \bar{z}) \, \phi_{\gamma}(z,\mu) &\nonumber \\
&=-\frac{Q^2}{Q^2 + s} \int_0^1 dz \, \theta \left (z-\frac{Q^2}{Q^2+s} \right ) \, \ln \left (\frac{z \, s-\bar{z} \, Q^2}{Q^2} \right ) \, \frac{d}{dz} \, \phi_{\gamma}(z,\mu) \,.
\end{flalign}
\begin{flalign}
&\frac{1}{\pi} \, \text{Im}_s \int_0^1 dz \, \frac{1}{z \, \bar{r}} \, \ln(z \, r+\bar{z}) \, \phi_{\gamma}(z,\mu) &\nonumber \\
& = -\frac{Q^2}{Q^2 + s} \int_0^1 \frac{dz}{z} \, \theta \left (z-\frac{Q^2}{Q^2+s} \right ) \, \phi_{\gamma}(z,\mu) \,.
\end{flalign}
\begin{flalign}
 &\frac{1}{\pi} \, \text{Im}_s \int_0^1 dz \, \frac{z\,r+\bar{z}}{z \, \bar{r}}  \, \ln (z \, r+\bar{z}) \,  \phi_{\gamma}(z,\mu) &\nonumber \\
& =\int_0^1 \frac{dz}{z} \, \theta \left (z-\frac{Q^2}{Q^2 + s} \right ) \, \frac{\bar{z} \, Q^2-z \,s}{Q^2+s} \, \phi_{\gamma}(z,\mu) \,.
\end{flalign}
\section{Leading Power Contributions of $\gamma^* \gamma \to f_2(1270)$}
\label{app:B}
We use the three helicity amplitudes for $\gamma^* \gamma \to f_2(1270)$ calculated in Ref.\cite{Braun:2016tsk} within the QCD collinear factorization framework, which incorporate higher-twist and partial radiative corrections. The explicit expressions are
\begin{align}
T_0^{\text{LP}} &= \langle f_q \rangle \int_0^1 \frac{du}{\bar{u}} \left[ 1 + \frac{\alpha_s}{4\pi} \mathbb{C}_q(u) \right] \phi_2(u) - \frac{\alpha_s}{4\pi} \frac{2}{3} f_g^S \int_0^1 du \, \mathbb{C}_g^S(u) \, \phi_g^S(u) \nonumber \\[0.5em]
&\quad + \frac{2m_{f_2}^2}{Q^2} \langle f_q \rangle \int_0^1 \frac{du}{\bar{u}} \left[ u \ln u \, \phi_2(u) - \frac{1}{8\bar{u}} \phi_4(u) \right], \\[0.5em]
T_1^{\text{LP}} &= 2 \langle f_q \rangle \int_0^1 \frac{du}{\bar{u}} \left[ g_v(u) - g_a(u) \right] \nonumber \\[0.5em]
&= 4 \langle f_q \rangle \int_0^1 \frac{du}{\bar{u}} \ln(u) \, \phi_2(u) + 2 \langle f_q \rangle \int \mathcal{D}\alpha \, \mathbb{C}_\Phi(\alpha) \left[ \Phi_3(\alpha) + \widetilde{\Phi}_3(\alpha) \right], \\[0.5em]
T_2^{\text{LP}} &= \frac{4m_{f_2}^2}{Q^2} \langle f_q \rangle \int_0^1 du \, \ln u \, g_v(u) + \frac{\alpha_s}{\pi} f_g^T \int_0^1 \frac{du}{\bar{u}} \left[ \frac{2}{3} + \frac{4}{9} \mathbb{C}_c(u) \right] \phi_g^T(u),
\end{align}
where the notation$\langle f_q \rangle$ stands for the sum of the light quark couplings weighted with the
electromagnetic charges
\begin{align}
\langle f_q \rangle = \frac{4}{9} f_u(\mu) + \frac{1}{9} f_d(\mu) + \frac{1}{9} f_s(\mu) = \frac{5\sqrt{2}}{18} f_q(\mu) + \frac{1}{9} f_s(\mu) \,.
\end{align}
We use $f_q$(1 GeV) $= 85(10) $ MeV, $\zeta_3 $(1 GeV)$= 0.15(8)$, $\omega_3$(1 GeV)$= -0.2(3)$, $\tilde{\omega}_3$(1 GeV)$= 0.06(1)$ as the default values for the present study\cite{Cheng:2010hn,Braun:2016tsk}. The coupling $f^S_g$ can be estimated from the radiative decay $\Upsilon(1S) \to \gamma f_2$~\cite{Fleming:2004hc}, in the numerical analysis we use the value $f^S_g$(1 GeV) = 45 MeV, and use $f^T_g$ (1 GeV) $\approx 20$MeV as a ballpark estimate. The function $\lambda(m_c^2/Q^2)$ takes into account suppression of the charm quark contribution in comparison to the light flavors:
\begin{align}
\lambda(x) =& 1 - 30x - 72x^2 + 24x(1 + 3x)\hat{\beta}\ln\left(\frac{\hat{\beta}+1}{\hat{\beta}-1}\right)- 6x\left(1 + 6x + 12x^2\right)\ln^2\left(\frac{\hat{\beta}+1}{\hat{\beta}-1}\right) \,,
\end{align}
where $\hat{\beta} = \sqrt{1 + 4x}$. The scale dependence of these parameters is as follows
\begin{align}
&f_{(8)} = \frac{1}{\sqrt{6}}\left(f_u + f_d - 2f_s\right) \,, \qquad f_{(1)} = \frac{1}{\sqrt{3}}\left(f_u + f_d + f_s\right) \,, \nonumber \\[0.5em]
&f_q(\mu) =  \sqrt{\frac{1}{3}} f_{(8)}(\mu) + \sqrt{\frac{2}{3}} f_{(1)}(\mu) \,, \qquad f_{s}(\mu)=-\sqrt{\frac{2}{3}}f_{(8)}(\mu)+\sqrt{\frac{1}{3}}f_{(1)}(\mu) \,,\nonumber \\[0.5em]
&f_{(8)}(\mu_0)= \sqrt{\frac{1}{3}} f_q(\mu_0) \,, \qquad f_{(1)}(\mu_0) = \sqrt{\frac{2}{3}} f_q(\mu_0) \,, \qquad f_{s}(\mu_{0}) = 0\,,\nonumber \\[0.5em]
&\left( \mu \frac{\partial}{\partial \mu} + \beta(g) \frac{\partial}{\partial g} \right)
\begin{pmatrix}
f_{(8)} \\
f_{(1)} \\
f_g^S
\end{pmatrix}
= \frac{\alpha_s}{2\pi}
\begin{pmatrix}
\frac{8}{3} C_F & 0 & 0 \\
0 & \frac{8}{3} C_F & -\frac{4}{3} \sqrt{n_f} \\
0 & -\frac{4}{3} \sqrt{n_f} C_F & \frac{2}{3} n_f
\end{pmatrix}
\begin{pmatrix}
f_{(8)} \\
f_{(1)} \\
f_g^S
\end{pmatrix}\,,
\end{align}
where
\begin{align}
&L = \frac{\alpha_s(\mu)}{\alpha_s(\mu_0)} \,, \qquad \beta_0 = \frac{11}{3}N_c - \frac{2}{3}n_f \,, \nonumber \\[0.5em]
&f_{(8)}(\mu) = L^{(\frac{8}{3} C_F)/\beta_0} f_{(8)}(\mu_0) \,, \nonumber \\[0.5em]
&f_{(1)}(\mu) = f_{(1)}(\mu_0) + \left[ L^{(\frac{8}{3} C_F + \frac{2}{3} n_f)/\beta_0} - 1 \right]
\left[ \frac{4 \, C_F}{4 \, C_F + n_f} f_{(1)}(\mu_0) - \frac{2\sqrt{n_f}}{4 \, C_F + n_f} f_g^S(\mu_0) \right] \,, \nonumber \\[0.5em]
&f_g^S(\mu) = f_g^S(\mu_0) - \left[ L^{(\frac{8}{3} C_F + \frac{2}{3} n_f)/\beta_0} - 1 \right]
\left[ \frac{2 \, C_F\sqrt{n_f}}{4 \, C_F + n_f} f_{(1)}(\mu_0) - \frac{n_f}{4C_F + n_f} f_g^S(\mu_0) \right] \,, \nonumber \\[0.5em]
&f_g^T(\mu) = L^{(\frac{7}{3} C_A + \frac{2}{3} n_f)/\beta_0} f_g^T(\mu_0) \,.
\end{align}
The asymptotic expression for the leading-twist DA of the $f_2(1270)$ meson
\begin{align}
\phi_2^{\mathrm{as}}(u)
=30u(1-u)(2u-1).
\end{align}
The asymptotic forms of the momentum fraction distribution of the two gluons in the $f_2$ meson with the same and the opposite helicity at large scales are
\begin{align}
\phi_g^{T,\text{as}}(u) = \phi_g^{S,\text{as}}(u) = 30u^2(1-u)^2.
\end{align}
The NLO quark and gluon coefficient functions for $T_0$ read
\begin{align}
\mathbb{C}_q(u) = C_F \left[ \ln^2 \bar{u} + 3 \ln u - 9 \right], \qquad \mathbb{C}_g^S(u) = \frac{2 \ln u}{u \bar{u}^2} \left[ u \ln u - 2u - 2 \right].
\end{align}
The coefficient function of the three-particle DAs to $T_1$ is given by
\begin{align}
\mathbb{C}_{\Phi}(\alpha) = \frac{1}{\alpha_2} \left[ \frac{1}{\alpha_1 \bar{\alpha}_1} + \frac{1}{\alpha_2} \left( \frac{\ln \alpha_1}{\bar{\alpha}_1} - \frac{\ln \bar{\alpha}_3}{\alpha_3} \right) + \frac{\ln \alpha_1}{\bar{\alpha}_1^2} \right] \, .
\end{align}
The c-quark contribution to the transversity gluon distribution is given by
\begin{align}
\mathbb{C}_c(u) =& 1 + \frac{2m_c^2}{Q^2} \biggl[ -\frac{\beta}{u\bar{u}} \ln\left( \frac{\beta+1}{\beta-1} \right) + \frac{\beta_{u}}{\bar{u}} \ln\left( \frac{\beta_{u} + 1}{\beta_{u} - 1} \right) + \frac{\beta_{\bar{u}}}{u} \ln\left( \frac{\beta_{\bar{u}} + 1}{\beta_{\bar{u}} - 1} \right) \nonumber \\[0.5em]
&+ \frac{1}{u\bar{u}} \left( \frac{1}{2} + \frac{m_c^2}{Q^2} \right) \left( \ln^2 \left( \frac{\beta+1}{\beta-1} \right) - \ln^2 \left( \frac{\beta_{u} + 1}{\beta_{u} - 1} \right) - \ln^2 \left( \frac{\beta_{\bar{u}} + 1}{\beta_{\bar{u}} - 1} \right) \right) \biggr] \,,
\end{align}
where
\begin{align}
\beta_u = \sqrt{1 + \frac{4m_c^2}{uQ^2}}, \qquad \beta \equiv \beta_1.
\end{align}
The asymptotic expression for $\phi_4(u)$ and $g_v(u)$ have the forms
\begin{align}
&\phi_4^{WW}(u) = 100 u^2 (1 - u)^2 (2u - 1), \nonumber \\[0.5em]
&g_v(u) = g_v^{WW}(u) - \left[ 10\zeta_3 - \frac{15}{8}(\omega_3 - \tilde{\omega}_3) \right] C_3^{1/2}(2u - 1) \, ,\nonumber  \\[0.5em]
&g_v^{WW}(u) = 3C_1^{1/2}(2u - 1) + 2C_3^{1/2}(2u - 1) \, .
\end{align}
The scale dependence of the flavor-nonsinglet twist-three couplings $\zeta_3 $, $\omega_3$, and $\tilde{\omega}_3$:
\begin{align}
&\zeta_3(\mu) = L^{3(C_A - C_F)/\beta_0} \zeta_3(\mu_0) \,, \nonumber \\[0.5em]
&\begin{pmatrix}
\tilde{\omega}_3 \\
\omega_3
\end{pmatrix}(\mu) = L^{\Gamma/\beta_0}
\begin{pmatrix}
\tilde{\omega}_3 \\
\omega_3
\end{pmatrix}(\mu_0) \,, \nonumber \\[0.5em]
&\Gamma = 
\begin{pmatrix}
\frac{13}{6} C_A - \frac{1}{12} C_F & \frac{7}{2} C_A - \frac{21}{4} C_F \\
\frac{1}{6} C_A - \frac{1}{4} C_F & \frac{25}{6} C_A - \frac{29}{12} C_F
\end{pmatrix}
=
\begin{pmatrix}
\frac{115}{18} & \frac{7}{2} \\
\frac{1}{6} & \frac{167}{18}
\end{pmatrix} \,.
\end{align}

The experimental results in \cite{Belle:2015oin} are presented for a different set of transition form factors
$F_i(Q^2)$ suggested in \cite{Schuler_1998}. To compare our predictions with the experimental data, we use the following simplified relations to translate between the two conventions:
\begin{align}
F_0(Q^2) &\simeq \sqrt{\frac{2}{3}} \left(1 + \frac{Q^2}{m_{f_2}^2}\right)^{-1} \left| \frac{T_0(Q^2)}{T_2(0)} \right| \,, \nonumber \\[0.5em]
F_1(Q^2) &\simeq \frac{\sqrt{Q^2/m_{f_2}^2}}{\left(1 + Q^2/m_{f_2}^2\right)^2} \left| \frac{T_1(Q^2)}{T_2(0)} \right| \,, \nonumber \\[0.5em]
F_2(Q^2) &\simeq \left(1 + \frac{Q^2}{m_{f_2}^2}\right)^{-1} \left| \frac{T_2(Q^2)}{T_2(0)} \right| \,.
\end{align}

\bibliographystyle{JHEP}
\bibliography{biblio.bib}

@article{BaBar:2008ozy,
    author = "Aubert, Bernard and others",
    collaboration = "BaBar",
    title = "{Observation and Polarization Measurements of $B^\pm \to \phi K_{1}^\pm$ and $B^\pm \to \phi K_{2}^{*\pm}$}",
    eprint = "0806.4419",
    archivePrefix = "arXiv",
    primaryClass = "hep-ex",
    reportNumber = "SLAC-PUB-13267, BABAR-PUB-08-021",
    doi = "10.1103/PhysRevLett.101.161801",
    journal = "Phys. Rev. Lett.",
    volume = "101",
    pages = "161801",
    year = "2008"
}

@article{BaBar:2008lan,
    author = "Aubert, Bernard and others",
    collaboration = "BaBar",
    title = "{Time-Dependent and Time-Integrated Angular Analysis of B -{\ensuremath{>}} phi Ks pi0 and B -{\ensuremath{>}} phi K+ pi-}",
    eprint = "0808.3586",
    archivePrefix = "arXiv",
    primaryClass = "hep-ex",
    reportNumber = "SLAC-PUB-13337, BABAR-PUB-08-036",
    doi = "10.1103/PhysRevD.78.092008",
    journal = "Phys. Rev. D",
    volume = "78",
    pages = "092008",
    year = "2008"
}

@article{Cheng:2010yd,
    author = "Cheng, Hai-Yang and Yang, Kwei-Chou",
    title = "{Charmless Hadronic B Decays into a Tensor Meson}",
    eprint = "1010.3309",
    archivePrefix = "arXiv",
    primaryClass = "hep-ph",
    reportNumber = "CYCU-HEP-10-16",
    doi = "10.1103/PhysRevD.83.034001",
    journal = "Phys. Rev. D",
    volume = "83",
    pages = "034001",
    year = "2011"
}

@article{Kim:2013cpa,
    author = "Kim, C. S. and Li, Run-Hui and Simanjuntak, Freddy and Zou, Z. T.",
    title = "{Charmless $B_{u,d,s} \to VT$ decays in perturbative QCD approach}",
    doi = "10.1103/PhysRevD.88.014031",
    journal = "Phys. Rev. D",
    volume = "88",
    number = "1",
    pages = "014031",
    year = "2013"
}

@article{Cheng:2010hn,
    author = "Cheng, Hai-Yang and Koike, Yuji and Yang, Kwei-Chou",
    title = "{Two-parton Light-cone Distribution Amplitudes of Tensor Mesons}",
    eprint = "1007.3541",
    archivePrefix = "arXiv",
    primaryClass = "hep-ph",
    reportNumber = "CYCU-HEP-10-08",
    doi = "10.1103/PhysRevD.82.054019",
    journal = "Phys. Rev. D",
    volume = "82",
    pages = "054019",
    year = "2010"
}

@article{Braun:2000cs,
    author = "Braun, Vladimir M. and Kivel, N.",
    title = "{Hard exclusive production of tensor mesons}",
    eprint = "hep-ph/0012220",
    archivePrefix = "arXiv",
    reportNumber = "TPR-00-23",
    doi = "10.1016/S0370-2693(01)00095-8",
    journal = "Phys. Lett. B",
    volume = "501",
    pages = "48--53",
    year = "2001"
}

@article{Braun:2016tsk,
    author = "Braun, V. M. and Kivel, N. and Strohmaier, M. and Vladimirov, A. A.",
    title = "{Electroproduction of tensor mesons in QCD}",
    eprint = "1603.09154",
    archivePrefix = "arXiv",
    primaryClass = "hep-ph",
    doi = "10.1007/JHEP06(2016)039",
    journal = "JHEP",
    volume = "06",
    pages = "039",
    year = "2016"
}

@article{Lebiedowicz:2020qnz,
    author = "Lebiedowicz, Piotr and Szczurek, Antoni",
    title = "{Production of $f_2(1270)$  meson in $pp$ collisions at the LHC via gluon-gluon fusion in the $k_t$-factorization approach}",
    eprint = "2007.12485",
    archivePrefix = "arXiv",
    primaryClass = "hep-ph",
    doi = "10.1016/j.physletb.2020.135816",
    journal = "Phys. Lett. B",
    volume = "810",
    pages = "135816",
    year = "2020"
}

@article{Giacosa:2005bw,
    author = "Giacosa, F. and Gutsche, Th. and Lyubovitskij, V. E. and Faessler, Amand",
    title = "{Decays of tensor mesons and the tensor glueball in an effective field approach}",
    eprint = "hep-ph/0511171",
    archivePrefix = "arXiv",
    doi = "10.1103/PhysRevD.72.114021",
    journal = "Phys. Rev. D",
    volume = "72",
    pages = "114021",
    year = "2005"
}

@article{Belle:2005rpz,
    author = "Garmash, A. and others",
    collaboration = "Belle",
    title = "{Evidence for large direct CP violation in B+- ---{\ensuremath{>}} rho(770)0K+- from analysis of the three-body charmless B+- ---{\ensuremath{>}}K+- pi+- pi-+ decay}",
    eprint = "hep-ex/0512066",
    archivePrefix = "arXiv",
    reportNumber = "BELLE-PREPRINT-2005-36",
    doi = "10.1103/PhysRevLett.96.251803",
    journal = "Phys. Rev. Lett.",
    volume = "96",
    pages = "251803",
    year = "2006"
}

@article{BaBar:2008lpx,
    author = "Aubert, Bernard and others",
    collaboration = "BaBar",
    title = "{Evidence for Direct CP Violation from Dalitz-plot analysis of $B^\pm \to K^\pm \pi^\mp \pi^\pm$}",
    eprint = "0803.4451",
    archivePrefix = "arXiv",
    primaryClass = "hep-ex",
    reportNumber = "BABAR-PUB-08-005, SLAC-PUB-13189",
    doi = "10.1103/PhysRevD.78.012004",
    journal = "Phys. Rev. D",
    volume = "78",
    pages = "012004",
    year = "2008"
}

@article{BESIII:2015rug,
    author = "Ablikim, M. and others",
    collaboration = "BESIII",
    title = "{Amplitude analysis of the $\pi^{0}\pi^{0}$~system produced in radiative $J/\psi$~decays}",
    eprint = "1506.00546",
    archivePrefix = "arXiv",
    primaryClass = "hep-ex",
    doi = "10.1103/PhysRevD.92.052003",
    journal = "Phys. Rev. D",
    volume = "92",
    number = "5",
    pages = "052003",
    year = "2015",
    note = "[Erratum: Phys.Rev.D 93, 039906 (2016)]"
}

@article{Belle:2015oin,
    author = "Masuda, M. and others",
    collaboration = "Belle",
    title = "{Study of $\pi^0$ pair production in single-tag two-photon collisions}",
    eprint = "1508.06757",
    archivePrefix = "arXiv",
    primaryClass = "hep-ex",
    reportNumber = "BELLE-PREPRINT-2015-15, KEK-PREPRINT-2015-24",
    doi = "10.1103/PhysRevD.93.032003",
    journal = "Phys. Rev. D",
    volume = "93",
    number = "3",
    pages = "032003",
    year = "2016"
}

@article{Schuler:1997yw,
    author = "Schuler, G. A. and Berends, Frits A. and van Gulik, R.",
    title = "{Meson photon transition form-factors and resonance cross-sections in e+ e- collisions}",
    eprint = "hep-ph/9710462",
    archivePrefix = "arXiv",
    reportNumber = "CERN-TH-97-294",
    doi = "10.1016/S0550-3213(98)00128-X",
    journal = "Nucl. Phys. B",
    volume = "523",
    pages = "423--438",
    year = "1998"
}

@article{Hoferichter:2020lap,
    author = "Hoferichter, Martin and Stoffer, Peter",
    title = "{Asymptotic behavior of meson transition form factors}",
    eprint = "2004.06127",
    archivePrefix = "arXiv",
    primaryClass = "hep-ph",
    reportNumber = "INT-PUB-20-015",
    doi = "10.1007/JHEP05(2020)159",
    journal = "JHEP",
    volume = "05",
    pages = "159",
    year = "2020"
}

@article{Wang:2017ijn,
    author = "Wang, Yu-Ming and Shen, Yue-Long",
    title = "{Subleading power corrections to the pion-photon transition form factor in QCD}",
    eprint = "1706.05680",
    archivePrefix = "arXiv",
    primaryClass = "hep-ph",
    reportNumber = "UWTHPH-2017-12",
    doi = "10.1007/JHEP12(2017)037",
    journal = "JHEP",
    volume = "12",
    pages = "037",
    year = "2017"
}

@article{Shen:2019zvh,
    author = {Shen, Yue-Long and Gao, Jing and L{\"u}, Cai-Dian and Miao, Yan},
    title = "{Power corrections to the pion transition form factor from higher-twist distribution amplitudes of a photon}",
    eprint = "1901.10259",
    archivePrefix = "arXiv",
    primaryClass = "hep-ph",
    doi = "10.1103/PhysRevD.99.096013",
    journal = "Phys. Rev. D",
    volume = "99",
    number = "9",
    pages = "096013",
    year = "2019"
}

@article{Stefanis:2020rnd,
    author = "Stefanis, N. G.",
    title = "{Pion-photon transition form factor in light cone sum rules and tests of asymptotics}",
    eprint = "2006.10576",
    archivePrefix = "arXiv",
    primaryClass = "hep-ph",
    reportNumber = "RUB-TPII-01/2020",
    doi = "10.1103/PhysRevD.102.034022",
    journal = "Phys. Rev. D",
    volume = "102",
    number = "3",
    pages = "034022",
    year = "2020"
}

@article{Agaev:2010aq,
    author = "Agaev, S. S. and Braun, V. M. and Offen, N. and Porkert, F. A.",
    title = "{Light Cone Sum Rules for the pi0-gamma*-gamma Form Factor Revisited}",
    eprint = "1012.4671",
    archivePrefix = "arXiv",
    primaryClass = "hep-ph",
    doi = "10.1103/PhysRevD.83.054020",
    journal = "Phys. Rev. D",
    volume = "83",
    pages = "054020",
    year = "2011"
}

@article{Mikhailov:2016klg,
    author = "Mikhailov, S. V. and Pimikov, A. V. and Stefanis, N. G.",
    title = "{Systematic estimation of theoretical uncertainties in the calculation of the pion-photon transition form factor using light-cone sum rules}",
    eprint = "1604.06391",
    archivePrefix = "arXiv",
    primaryClass = "hep-ph",
    reportNumber = "RUB-TPII-01-2016",
    doi = "10.1103/PhysRevD.93.114018",
    journal = "Phys. Rev. D",
    volume = "93",
    number = "11",
    pages = "114018",
    year = "2016"
}

@article{Agaev:2014wna,
    author = {Agaev, S. S. and Braun, V. M. and Offen, N. and Porkert, F. A. and Sch{\"a}fer, A.},
    title = "{Transition form factors $\gamma^*\gamma\to\eta$ and $\gamma^*\gamma\to\eta'$ in QCD}",
    eprint = "1409.4311",
    archivePrefix = "arXiv",
    primaryClass = "hep-ph",
    doi = "10.1103/PhysRevD.90.074019",
    journal = "Phys. Rev. D",
    volume = "90",
    number = "7",
    pages = "074019",
    year = "2014"
}

@article{Braun:2025xpc,
    author = "Braun, Vladimir M. and Chetyrkin, Konstantin G. and Manashov, Alexander N.",
    title = "{{\ensuremath{\gamma}}*{\textrightarrow}{\ensuremath{\eta}}{\ensuremath{\gamma}} and {\ensuremath{\gamma}}*{\textrightarrow}{\ensuremath{\eta}}'{\ensuremath{\gamma}} form factors to NNLO accuracy in perturbative QCD}",
    eprint = "2510.15643",
    archivePrefix = "arXiv",
    primaryClass = "hep-ph",
    reportNumber = "DESY-25-141",
    doi = "10.1103/47v7-ss7l",
    journal = "Phys. Rev. D",
    volume = "113",
    number = "5",
    pages = "056026",
    year = "2026"
}

@article{Sun:2010nv,
    author = "Sun, Yan-Jun and Li, Zuo-Hong and Huang, Tao",
    title = "{$B_{(s)}\to S$ transitions in the light cone sum rules with the chiral current}",
    eprint = "1011.3901",
    archivePrefix = "arXiv",
    primaryClass = "hep-ph",
    doi = "10.1103/PhysRevD.83.025024",
    journal = "Phys. Rev. D",
    volume = "83",
    pages = "025024",
    year = "2011"
}

@article{Zhong:2011jf,
    author = "Zhong, Tao and Wu, Xing-Gang and Zhang, Jia-Wei and Tang, Yun-Qing and Fang, Zhen-Yun",
    title = "{New results on Pionic Twist-3 Distribution Amplitudes within the QCD Sum Rules}",
    eprint = "1101.3592",
    archivePrefix = "arXiv",
    primaryClass = "hep-ph",
    doi = "10.1103/PhysRevD.83.036002",
    journal = "Phys. Rev. D",
    volume = "83",
    pages = "036002",
    year = "2011"
}

@article{Wang:2017jow,
    author = {Wang, Yu-Ming and Wei, Yan-Bing and Shen, Yue-Long and L{\"u}, Cai-Dian},
    title = "{Perturbative corrections to B {\textrightarrow} D form factors in QCD}",
    eprint = "1701.06810",
    archivePrefix = "arXiv",
    primaryClass = "hep-ph",
    reportNumber = "UWTHPH-2016-29",
    doi = "10.1007/JHEP06(2017)062",
    journal = "JHEP",
    volume = "06",
    pages = "062",
    year = "2017"
}

@article{Wang:2018wfj,
    author = "Wang, Yu-Ming and Shen, Yue-Long",
    title = "{Subleading-power corrections to the radiative leptonic $B \to \gamma \ell \nu$ decay in QCD}",
    eprint = "1803.06667",
    archivePrefix = "arXiv",
    primaryClass = "hep-ph",
    doi = "10.1007/JHEP05(2018)184",
    journal = "JHEP",
    volume = "05",
    pages = "184",
    year = "2018"
}

@article{Emmerich:2018rug,
    author = {Emmerich, M. and Strohmaier, M. and Sch{\"a}fer, A.},
    title = "{B $\rightarrow f_2(1270)$ form factors with light-cone sum rules}",
    eprint = "1804.02953",
    archivePrefix = "arXiv",
    primaryClass = "hep-ph",
    doi = "10.1103/PhysRevD.98.014008",
    journal = "Phys. Rev. D",
    volume = "98",
    number = "1",
    pages = "014008",
    year = "2018"
}

@article{Gao:2019lta,
    author = {Gao, Jing and L{\"u}, Cai-Dian and Shen, Yue-Long and Wang, Yu-Ming and Wei, Yan-Bing},
    title = "{Precision calculations of $B \to V$ form factors from soft-collinear effective theory sum rules on the light-cone}",
    eprint = "1907.11092",
    archivePrefix = "arXiv",
    primaryClass = "hep-ph",
    doi = "10.1103/PhysRevD.101.074035",
    journal = "Phys. Rev. D",
    volume = "101",
    number = "7",
    pages = "074035",
    year = "2020"
}

@article{Beneke:2020fot,
    author = "Beneke, Martin and Bobeth, Christoph and Wang, Yu-Ming",
    title = "{$B_{d,s}\to\gamma\ell\bar{\ell}$ decay with an energetic photon}",
    eprint = "2008.12494",
    archivePrefix = "arXiv",
    primaryClass = "hep-ph",
    reportNumber = "TUM-HEP-1280/20",
    doi = "10.1007/JHEP12(2020)148",
    journal = "JHEP",
    volume = "12",
    pages = "148",
    year = "2020"
}

@article{Li:2020rcg,
    author = {Li, Hua-Dong and L{\"u}, Cai-Dian and Wang, Chao and Wang, Yu-Ming and Wei, Yan-Bing},
    title = "{QCD calculations of radiative heavy meson decays with subleading power corrections}",
    eprint = "2002.03825",
    archivePrefix = "arXiv",
    primaryClass = "hep-ph",
    doi = "10.1007/JHEP04(2020)023",
    journal = "JHEP",
    volume = "04",
    pages = "023",
    year = "2020"
}

@article{Gao:2021sav,
    author = "Gao, Jing and Huber, Tobias and Ji, Yao and Wang, Chao and Wang, Yu-Ming and Wei, Yan-Bing",
    title = "{B {\textrightarrow} D{\ensuremath{\ell}}{\ensuremath{\nu}}$_{ℓ}$ form factors beyond leading power and extraction of $|V_{cb}|$ and R(D)}",
    eprint = "2112.12674",
    archivePrefix = "arXiv",
    primaryClass = "hep-ph",
    reportNumber = "TUM-HEP-1331/21, SI-HEP-2021-37, P3H-21-103",
    doi = "10.1007/JHEP05(2022)024",
    journal = "JHEP",
    volume = "05",
    pages = "024",
    year = "2022"
}

@article{Cui:2022zwm,
    author = "Cui, Bo-Yan and Huang, Yong-Kang and Shen, Yue-Long and Wang, Chao and Wang, Yu-Ming",
    title = "{Precision calculations of B$_{d,s}$ {\textrightarrow} {\ensuremath{\pi}}, K decay form factors in soft-collinear effective theory}",
    eprint = "2212.11624",
    archivePrefix = "arXiv",
    primaryClass = "hep-ph",
    doi = "10.1007/JHEP03(2023)140",
    journal = "JHEP",
    volume = "03",
    pages = "140",
    year = "2023"
}

@article{Huang:2025jsa,
    author = {Huang, Yong-Kang and Li, Dong-Hao and L{\"u}, Cai-Dian and Shi, Bo-Xuan and Yu, Hui-Xin},
    title = "{Next-to-Next-to-Leading-Order Corrections to the $B \to \pi$ Form Factors from Light-Cone Sum Rules}",
    eprint = "2512.18866",
    archivePrefix = "arXiv",
    primaryClass = "hep-ph",
    month = "12",
    year = "2025"
}

@article{Di:2025hdu,
    author = "Di, Fang-Zhou and Liu, Yu-Hao",
    title = "{Next-to-leading-order QCD calculations of B{\textrightarrow}A form factors with higher-twist corrections}",
    eprint = "2504.13649",
    archivePrefix = "arXiv",
    primaryClass = "hep-ph",
    doi = "10.1103/k5qy-rc5j",
    journal = "Phys. Rev. D",
    volume = "112",
    number = "11",
    pages = "116003",
    year = "2025"
}

@article{Gao:2024vql,
    author = "Gao, Jing and Mei{\ss}ner, Ulf-G. and Shen, Yue-Long and Li, Dong-Hao",
    title = "{Precision calculations of B{\textrightarrow}K* form factors from SCET sum rules beyond leading-power contributions}",
    eprint = "2412.13084",
    archivePrefix = "arXiv",
    primaryClass = "hep-ph",
    doi = "10.1103/yvjd-2ymn",
    journal = "Phys. Rev. D",
    volume = "112",
    number = "1",
    pages = "014032",
    year = "2025"
}

@article{Wang_2011,
   title={B to tensor meson form factors in the perturbative QCD approach},
   volume={83},
   ISSN={1550-2368},
   url={http://dx.doi.org/10.1103/PhysRevD.83.014008},
   DOI={10.1103/physrevd.83.014008},
   number={1},
   journal={Physical Review D},
   publisher={American Physical Society (APS)},
   author={Wang, Wei},
   year={2011},
   month=jan }

@article{Belitsky:1999gu,
    author = "Belitsky, Andrei V. and Mueller, Dieter and Freund, A.",
    title = "{Reconstruction of nonforward evolution kernels}",
    eprint = "hep-ph/9904477",
    archivePrefix = "arXiv",
    doi = "10.1016/S0370-2693(99)00837-0",
    journal = "Phys. Lett. B",
    volume = "461",
    pages = "270--279",
    year = "1999"
}

@article{Belitsky:2000yn,
    author = "Belitsky, Andrei V. and Freund, A. and Mueller, Dieter",
    title = "{NLO evolution kernels for skewed transversity distributions}",
    eprint = "hep-ph/0008005",
    archivePrefix = "arXiv",
    doi = "10.1016/S0370-2693(00)01129-1",
    journal = "Phys. Lett. B",
    volume = "493",
    pages = "341--349",
    year = "2000"
}

@article{Ball_2003,
   title={Photon distribution amplitudes in QCD},
   volume={649},
   ISSN={0550-3213},
   url={http://dx.doi.org/10.1016/S0550-3213(02)01017-9},
   DOI={10.1016/s0550-3213(02)01017-9},
   number={1–2},
   journal={Nuclear Physics B},
   publisher={Elsevier BV},
   author={Ball, Patricia and Braun, V.M. and Kivel, N.},
   year={2003},
   month=jan, pages={263–296} }

@article{Mikhailov:2008my,
    author = "Mikhailov, S. V. and Vladimirov, A. A.",
    title = "{ERBL and DGLAP kernels for transversity distributions. Two-loop calculations in covariant gauge}",
    eprint = "0810.1647",
    archivePrefix = "arXiv",
    primaryClass = "hep-ph",
    doi = "10.1016/j.physletb.2008.11.051",
    journal = "Phys. Lett. B",
    volume = "671",
    pages = "111--118",
    year = "2009"
}

@article{ParticleDataGroup:2014cgo,
    author = "Olive, K. A. and others",
    collaboration = "Particle Data Group",
    title = "{Review of Particle Physics}",
    doi = "10.1088/1674-1137/38/9/090001",
    journal = "Chin. Phys. C",
    volume = "38",
    pages = "090001",
    year = "2014"
}

@article{FlavourLatticeAveragingGroupFLAG:2021npn,
    author = "Aoki, Y. and others",
    collaboration = "Flavour Lattice Averaging Group (FLAG)",
    title = "{FLAG Review 2021}",
    eprint = "2111.09849",
    archivePrefix = "arXiv",
    primaryClass = "hep-lat",
    reportNumber = "CERN-TH-2021-191, JLAB-THY-21-3528, FERMILAB-PUB-21-620-SCD-T",
    doi = "10.1140/epjc/s10052-022-10536-1",
    journal = "Eur. Phys. J. C",
    volume = "82",
    number = "10",
    pages = "869",
    year = "2022"
}

@article{Schuler_1998,
   title={Meson-photon transition form factors and resonance cross-sections in $e^+e^-$ collisions},
   volume={523},
   ISSN={0550-3213},
   url={http://dx.doi.org/10.1016/S0550-3213(98)00128-X},
   DOI={10.1016/s0550-3213(98)00128-x},
   number={3},
   journal={Nuclear Physics B},
   publisher={Elsevier BV},
   author={Schuler, G.A. and Berends, F.A. and van Gulik, R.},
   year={1998},
   month=jul, pages={423–438} }

@article{Wang_2017,
   title={Subleading power corrections to the pion-photon transition form factor in QCD},
   volume={2017},
   ISSN={1029-8479},
   url={http://dx.doi.org/10.1007/JHEP12(2017)037},
   DOI={10.1007/jhep12(2017)037},
   number={12},
   journal={Journal of High Energy Physics},
   publisher={Springer Science and Business Media LLC},
   author={Wang, Yu-Ming and Shen, Yue-Long},
   year={2017},
   month=dec }

@article{Aliev:1981ju,
    author = "Aliev, T. M. and Shifman, Mikhail A.",
    title = "{Old Tensor Mesons in {QCD} Sum Rules}",
    reportNumber = "ITEP-133-1981",
    doi = "10.1016/0370-2693(82)91078-4",
    journal = "Phys. Lett. B",
    volume = "112",
    pages = "401--405",
    year = "1982"
}

@article{Aliev:1982ab,
    author = "Aliev, T. M. and Shifman, Mikhail A.",
    title = "{QCD SUM RULES AND TENSOR MESONS. (IN RUSSIAN)}",
    journal = "Sov. J. Nucl. Phys.",
    volume = "36",
    pages = "891",
    year = "1982"
}

@article{Cheng:2005nb,
    author = "Cheng, Hai-Yang and Chua, Chun-Khiang and Yang, Kwei-Chou",
    title = "{Charmless hadronic B decays involving scalar mesons: Implications to the nature of light scalar mesons}",
    eprint = "hep-ph/0508104",
    archivePrefix = "arXiv",
    doi = "10.1103/PhysRevD.73.014017",
    journal = "Phys. Rev. D",
    volume = "73",
    pages = "014017",
    year = "2006"
}

@article{Lepage:1979zb,
    author = "Lepage, G. Peter and Brodsky, Stanley J.",
    title = "{Exclusive Processes in Quantum Chromodynamics: Evolution Equations for Hadronic Wave Functions and the Form-Factors of Mesons}",
    reportNumber = "SLAC-PUB-2343",
    doi = "10.1016/0370-2693(79)90554-9",
    journal = "Phys. Lett. B",
    volume = "87",
    pages = "359--365",
    year = "1979"
}

@article{Shifman:1980dk,
    author = "Shifman, Mikhail A. and Vysotsky, Michael I.",
    title = "{FORM-FACTORS OF HEAVY MESONS IN QCD}",
    reportNumber = "ITEP-147-1980",
    doi = "10.1016/0550-3213(81)90023-7",
    journal = "Nucl. Phys. B",
    volume = "186",
    pages = "475--518",
    year = "1981"
}

@article{Ball:2002ps,
    author = "Ball, Patricia and Braun, V. M. and Kivel, N.",
    title = "{Photon distribution amplitudes in QCD}",
    eprint = "hep-ph/0207307",
    archivePrefix = "arXiv",
    reportNumber = "IPPP-02-40, DCPT-02-80",
    doi = "10.1016/S0550-3213(02)01017-9",
    journal = "Nucl. Phys. B",
    volume = "649",
    pages = "263--296",
    year = "2003"
}

@article{Ball:2006wn,
    author = "Ball, Patricia and Braun, V. M. and Lenz, A.",
    title = "{Higher-twist distribution amplitudes of the K meson in QCD}",
    eprint = "hep-ph/0603063",
    archivePrefix = "arXiv",
    reportNumber = "IPPP-06-01, DCPT-06-02, CERN-PH-TH-2006-026",
    doi = "10.1088/1126-6708/2006/05/004",
    journal = "JHEP",
    volume = "05",
    pages = "004",
    year = "2006"
}

@article{Colangelo:2000dp,
    author = "Colangelo, Pietro and Khodjamirian, Alexander",
    editor = "Shifman, M. and Ioffe, Boris",
    title = "{QCD sum rules, a modern perspective}",
    eprint = "hep-ph/0010175",
    archivePrefix = "arXiv",
    reportNumber = "CERN-TH-2000-296, BARI-TH-2000-394",
    doi = "10.1142/9789812810458_0033",
    pages = "1495--1576",
    month = "10",
    year = "2000"
}

@inproceedings{Braun:1997kw,
    author = "Braun, Vladimir M.",
    title = "{Light cone sum rules}",
    booktitle = "{4th International Workshop on Progress in Heavy Quark Physics}",
    eprint = "hep-ph/9801222",
    archivePrefix = "arXiv",
    reportNumber = "NORDITA-98-1-P",
    pages = "105--118",
    month = "9",
    year = "1997"
}

@article{Khodjamirian:2023wol,
    author = "Khodjamirian, Alexander and Meli{\'c}, Bla{\v{z}}enka and Wang, Yu-Ming",
    title = "{A guide to the QCD light-cone sum rules for $b$-quark decays}",
    eprint = "2311.08700",
    archivePrefix = "arXiv",
    primaryClass = "hep-ph",
    reportNumber = "SI-HEP-2023-25, P3H-23-087; RBI-ThPhys-2023-37",
    doi = "10.1140/epjs/s11734-023-01046-6",
    journal = "Eur. Phys. J. ST",
    volume = "233",
    number = "2",
    pages = "271--298",
    year = "2024"
}

@article{Beneke:1997zp,
    author = "Beneke, M. and Smirnov, Vladimir A.",
    title = "{Asymptotic expansion of Feynman integrals near threshold}",
    eprint = "hep-ph/9711391",
    archivePrefix = "arXiv",
    reportNumber = "CERN-TH-97-315",
    doi = "10.1016/S0550-3213(98)00138-2",
    journal = "Nucl. Phys. B",
    volume = "522",
    pages = "321--344",
    year = "1998"
}

@article{Fleming:2004hc,
    author = "Fleming, Sean and Lee, Christopher and Leibovich, Adam K.",
    title = "{Exclusive radiative decays of Upsilon in SCET}",
    eprint = "hep-ph/0411180",
    archivePrefix = "arXiv",
    reportNumber = "CALT-68-2530",
    doi = "10.1103/PhysRevD.71.074002",
    journal = "Phys. Rev. D",
    volume = "71",
    pages = "074002",
    year = "2005"
}

\end{document}